\documentclass{aastex631}

\newcommand{\DM}{{\rm DM}}

\usepackage{CJK}
\usepackage{physics}
\usepackage{booktabs}
\usepackage{hyperref}

\newcommand{\LowDiffCEH}{{70.4_{-2.5}^{+2.6}}}
\newcommand{\LowDiffCEOmega}{{0.6_{-0.3}^{+0.3}}}
\newcommand{\LowDiffCEw}{{-0.8_{-0.7}^{+0.6}}}
\newcommand{\LowDiffLVKH}{{73.6_{-3.5}^{+3.7}}}
\newcommand{\LowDiffLVKOmega}{{0.4_{-0.3}^{+0.4}}}
\newcommand{\LowDiffLVKw}{{-1.2_{-0.6}^{+0.7}}}

\newcommand{\LowExtCEH}{{71.1_{-4.4}^{+4.6}}}
\newcommand{\LowExtCEOmega}{{0.6_{-0.4}^{+0.3}}}
\newcommand{\LowExtCEw}{{-0.8_{-0.7}^{+0.6}}}
\newcommand{\LowExtCEmu}{{147.6_{-18.3}^{+18.7}}}
\newcommand{\LowExtCEsigma}{{0.6_{-0.1}^{+0.1}}}
\newcommand{\LowExtLVKH}{{79.7_{-7.1}^{+7.7}}}
\newcommand{\LowExtLVKOmega}{{0.4_{-0.3}^{+0.4}}}
\newcommand{\LowExtLVKw}{{-1.1_{-0.6}^{+0.7}}}
\newcommand{\LowExtLVKmu}{{143.3_{-19.9}^{+18.8}}}
\newcommand{\LowExtLVKsigma}{{0.5_{-0.1}^{+0.1}}}

\newcommand{\HighDiffMqH}{{68.9_{-1.5}^{+1.4}}}
\newcommand{\HighDiffMqOmega}{{0.3_{-0.2}^{+0.1}}}
\newcommand{\HighDiffMqw}{{-0.9_{-0.7}^{+0.4}}}
\newcommand{\HighDiffLnH}{{68.1_{-1.5}^{+1.6}}}
\newcommand{\HighDiffLnOmega}{{0.4_{-0.2}^{+0.1}}}
\newcommand{\HighDiffLnw}{{-1.0_{-0.7}^{+0.5}}}

\newcommand{\HighExtMqH}{{68.7_{-2.0}^{+2.3}}}
\newcommand{\HighExtMqOmega}{{0.4_{-0.2}^{+0.1}}}
\newcommand{\HighExtMqw}{{-1.1_{-0.7}^{+0.6}}}
\newcommand{\HighExtMqmu}{{287.4_{-91.4}^{+74.1}}}
\newcommand{\HighExtMqsigma}{{0.9_{-0.2}^{+0.3}}}
\newcommand{\HighExtLnH}{{68.0_{-2.2}^{+2.2}}}
\newcommand{\HighExtLnOmega}{{0.4_{-0.2}^{+0.1}}}
\newcommand{\HighExtLnw}{{-0.9_{-0.7}^{+0.5}}}
\newcommand{\HighExtLnmu}{{204.4_{-89.0}^{+98.3}}}
\newcommand{\HighExtLnsigma}{{0.7_{-0.3}^{+0.4}}}

\graphicspath{{./}{Figures/}}

\begin{document}
\begin{CJK}{UTF8}{gbsn}

\title{Cosmological Constraints from Gravitational Wave--Fast Radio Burst Associations without Redshift Measurements for LIGO-Virgo and Cosmic Explorer}

\author[0000-0001-8114-3094]{Jiaming Zhuge}\thanks{jiaming.zhuge@unlv.edu}
\affiliation{Nevada Center for Astrophysics, University of Nevada, Las Vegas, NV 89154, USA}
\affiliation{Department of Physics and Astronomy, University of Nevada, Las Vegas, NV 89154, USA}

\author[0000-0001-6677-949X]{Marios Kalomenopoulos}\thanks{marios.kalomenopoulos@gmail.com}
\affiliation{Nevada Center for Astrophysics, University of Nevada, Las Vegas, NV 89154, USA}
\affiliation{Department of Physics and Astronomy, University of Nevada, Las Vegas, NV 89154, USA}

\author[0000-0001-8040-9807]{Carl-Johan Haster}\thanks{carl.haster@unlv.edu}
\affiliation{Nevada Center for Astrophysics, University of Nevada, Las Vegas, NV 89154, USA}
\affiliation{Department of Physics and Astronomy, University of Nevada, Las Vegas, NV 89154, USA}

\author[0000-0002-9725-2524]{Bing Zhang}\thanks{bing.zhang@unlv.edu}
\affiliation{Nevada Center for Astrophysics, University of Nevada, Las Vegas, NV 89154, USA}
\affiliation{Department of Physics and Astronomy, University of Nevada, Las Vegas, NV 89154, USA}

\begin{abstract}
The potential association between gravitational waves (GWs) and fast radio bursts (FRBs) offers a unique multi-messenger probe for cosmology. In this paper, we develop a redshift-independent framework to constrain cosmological parameters using the luminosity distance - dispersion measure relation, accounting for realistic astrophysical uncertainties. We perform a comprehensive comparative analysis across different GWs detector sensitivities and modeling assumptions. Specifically, we investigate the performance of the current LIGO-Virgo (LV) network (at $z < 0.2$) versus the future Cosmic Explorer (CE). Our study further evaluates the impact of different dispersion measure (DM) distributions---specifically the corrected Macquart's PDF (Zhuge+2025) and the log-normal distribution---and explores the influence of including or excluding host galaxy DM contributions. Using realistic simulated observations, we find that while the current LV network lacks the precision to provide meaningful constraints, CE will enable high-precision cosmology. Even without spectroscopic redshifts, CE observations can effectively break parameter degeneracies and robustly constrain both cosmology and host galaxy parameters. These results highlight the necessity of next-generation detectors. 
\end{abstract}

\keywords{Gravitational waves (678) --- Gravitational wave astronomy (675) --- Radio bursts (1339) --- Radio transient sources (2008) --- Cosmological parameters (339)}

\section{Introduction} \label{sec:intro}

The modern concordance model of the Universe, $\Lambda$CDM is largely successful when confronted with observations of different scales \citep{LCDM_DE_Carroll_2001, Peebles_Ratra_2003, LCDM_Bull_et_al_2016, dodelson2024modern}. However, in recent years, observations of increased precision have presented new challenges \citep{Perivolaropoulos_Skara_2022, Peebles_LCDM_overview_2025}. One of the most famous puzzles is the ``Hubble Tension'', i.e. the statistically significant differences on the Hubble-Lema\^{i}tre parameter $H_0$ \citep{Hubble_1929, Lemaitre_1931} obtained by different experiments. ``Early Universe'' experiments constrain its value to $H_0 \simeq 67$ km/s/Mpc \citep{Planck_Cosmo_param_2018}, while ``late Universe'' observations measure $H_0$ as $\simeq 73$ km/s/Mpc \citep{SN_Ho_2019, SN_Ho_2021}, with statistical uncertainties small enough for the $H_0$ posteriors to not overlap at more than $3\sigma$. In recent years, the statistical uncertainties can even reach a $\sim 7\sigma$ difference between  ``early Universe'' and ``late Universe'' measurement, with $H_0 = 67.19 \pm 0.38$ km/s/Mpc \citep{2026PhRvDH06719} and $H_0 = 73.50\pm 0.81$ km/s/Mpc \citep{2026AnAH07350} respectively\footnote{Uncertainties correspond to $68$\% confidence intervals.}. In addition, the latest data release of Dark Energy Spectroscopic Instrument (DESI) suggested possible hints for a more complex dark energy (DE) component, inconsistent with a cosmological constant \citep{DESI_2025}. A plethora of theoretical solutions have been proposed \citep{Di_Valentino_et_al_2021_HoSolutions}, but novel, independent probes, with different systematics are required to be able to resolve the current open questions.

In recent years two new astronomical signals, Fast Radio Bursts (FRBs) \citep{Lorimer_burst_2007, Petroff_FRB_review_2019, Petroff_FRB_review_2022, Physics_FRBs_Bing_2023} and Gravitational Waves (GWs) \citep{FirstDetection, First_Run_2016, GW170817_H0_measurement_2017, GWTC-1_2019, GWTC-2_2021, GWTC-3_2023, LIGOScientific:2021usb, GWTC4_Catalogue_2025}, have been used in cosmological studies \citep{Glowacki_Lee_FRB_cosmo_Review_2024, Wu_Wang_FRB_cosmo_review_2024, Chen_et_al_2024, Palmese_Mastrogiovanni_2025_Ho_cosmo_review, Poggiani_2025_Ho_cosmo_review, LIGOScientific:2025jau, 2026SCPMA_Jin_StandardSirens}, which add new possibilities to resolve the cosmological questions. 

FRBs are transient, radio signals of largely unknown sources\footnote{For one case, FRB 20200428D, its source was associated with a galactic magnetar \citep{FRB_magnetar_CHIME_2020, FRB_magnetar_Bochenek_2020}.} \citep{FRB_living_theory_Platts_et_al_2019}, with the majority of them of cosmological origin \citep{Cordes_Chatterjee_FRB_review_2019, Petroff_FRB_review_2019, Petroff_FRB_review_2022, Physics_FRBs_Bing_2023}. They present a frequency dispersion, which depends on the electron number density in the line of sight and their distance from Earth. As a result, this feature makes them an interesting probe for cosmological constraints. In most cases, a counterpart is needed to provide a redshift estimate \citep{Glowacki_Lee_FRB_cosmo_Review_2024}. Over the past few years, the number of well-localized FRBs has increased and different cosmological parameters have been inferred in the literature, from the Hubble parameter $H_0$, to different cosmological densities $\Omega_i$, to the ``missing baryons'' of the Universe \citep[e.g.][]{Macquart_relation_2020,FRBcosmo_localised_James_et_al_2022, FRBcosmo_localised_Hagstotz_et_al_2022, FRBcosmo_localised_Fortunato_et_al_2023, FRBcosmo_localised_Kalita_et_al_2025, FRBcosmo_localised_Zhuge_et_al_2025, FRBcosmo_localised_Gao_et_al_2025, Connor_2025NatAs}.

In parallel, GWs from compact binary mergers have been used to constrain $H_0$. The single observation with the most constraining measurement comes from GW170817, a binary neutron star (NS) merger \citep{GW170817_detection_2017}, where an electromagnetic (EM) counterpart was detected and redshift was acquired \citep{MMA_GW170817_2017}. The Hubble parameter was constrained to $H_0 = 70.0^{+12.0}_{-8.0}$ km s$^{-1}$ Mpc$^{-1}$ \citep{GW170817_H0_measurement_2017}. This is still the only GWs event with a confirmed EM counterpart so far\footnote{GW190521 is another GW event - a potentially dynamically formed BBH, associated with an EM flare in an AGN disk \citep{GW190521_LVK_2020, GW190521_LVK_properties_2020, GW190521_Romero-Shaw_et_al_2020, GW190521_Flare_Graham_et_al_2020} - with cosmological applications as a standard-siren \citep{GW190521_Mukherjee_et_al_2020, GW190521_Gayathri_et_al_2020, GW190521_Bustillo_et_al_2021, Chen:2020gek}.}. In the absence of an EM source to provide redshift information, statistical methods can be used to derive cosmological constraints: ``dark sirens'' rely on information from galaxy catalogues in the GWs sky localisation region \citep{Schutz_1986, Holz_Hughes_2005, Dark_Sirens_Del_Pozzo_2012, Soares-Santos_et_al_2019, DarkSirens_Gray_et_al_2020, DarkSirens_Finke_et_al_2021, DarkSirens_Gair_et_al_2023}, ``spectral sirens'' rely on features of the binary black hole (BH) population \citep{SpectralSirens_Chernoff_Finn_1993, SpectralSirens_Taylor_et_al_2012, SpectralSirens_Ezquiaga_Holz_2022, Spectral_Sirens_Hernandez_Ray_2024, Spectral_Sirens_Mali_Essick_2025, SpectralSirens_Farah_et_al_2025}, while other methods rely on statistical cross-correlations of the GWs and galaxies' distributions \citep{GWsCorrelations_Mukherjee_et_al_2021, GWsCorrelations_Mukherjee_et_al_2024, GWsCorrelations_Ferri_et_al_2025} and phase information of the NS waveforms \citep{Phase_effects_Nishizawa_2012, Phase_Messenger_Read_2012, Phase_Messenger_et_al_2014}.

There are a number of proposals suggesting that GWs and FRBs signals are associated \citep{FRB_GW_all_model_arxiv2026}, particularly in the context of compact binary mergers \citep{FRBGW_BNs_2001MNRAS.322..695H, FRBGW_CBCs_2012ApJ...757L...3L,  FRBGW_BNs_2012ApJ...755...80P, FRBGW_BNs_2014ApJ...780L..21Z, FRBGW_BHNS_2015ApJ...814L..20M, FRBGW_BNs_2016ApJ...822L...7W, FRBGW_BH_2018PhRvD..98l3002L, FRBGW_BNs_2019ApJ...886..110M, FRBGW_BNs_2020ApJ...891...72W,  FRBGW_BHNS_2019ApJ...873L..13D, FRBGW_BHNS_2023ApJ...956L..33M, 2023ApJ_Ruinan}. See \cite{MMA_FRBS_Bing_2024} for a review of GW-FRB association models. At the time of writing no statistically significant association has been made between a GWs signal and an FRB or any other short radio burst \citep{LVK_GWs_Radio_bursts_2016, LVK_GWs_Magnetar_Assoc_2019, Wang_Nitz_GWs_FRB_Assoc_2022, LVK_CHIME_GWs_FRB_Assoc_2023, LVK_GWs_Magnetar_Assoc_2024}. If one of these proposed scenarios is correct, multimessenger analysis using current and future observatories, such as the Deep Synoptic Array (DSA) \citep{DSA_obs_2019}, the Canadian Hydrogen Intensity Mapping Experiment (CHIME) \citep{CHIME_PathFinder_2014, CHIME_FRB_System_2018}, the Laser Interferometer Gravitational-Wave Observatory (LIGO) \citep{Advanced_LIGO_2015}, Advanced Virgo \citep{Advanced_VIRGO_2015}, Kamioka Gravitational Wave Detector (KAGRA) \citep{KAGRA_Interferometer_2013}, and Cosmic Explorer (CE) \citep{CE_sensitivity_2021} could capture these GWs-FRB association events \citep{BNS_MMA_followUp_LVK_et_al_2018}. Such events provide a novel method for constraining cosmological parameters. The strength of this method lies in the fact that it does not require redshift information for the sources; however, it assumes that the FRB and GW signals originate from the same source. \citet{Wei_et_al_FRB_GWs_Ass_2018} first noted this possibility and constructed an $H_0$-free parameter, $D_L \times \DM$ to obtain constraints on other cosmological parameters. Furthermore, \citet{Jiguo_Zhang_GW_FRB} analyzed FRBs and GWs as independent events in an effort to remove any $H_0$ biases from the FRBs' baryon content determination. However, both of these studies assumed highly localized FRBs, i.e. assumed that the redshift of the sources was known. In contrast, \citet{Jahns-Schindler_Spitler_FRBGWsCosmo_2025} combined FRBs DM and GWs luminosity distance $D_L$ measurements of associated events to break the $H_0$ and density of matter $\Omega_b f_d$ degeneracy from GWs as a proxy for redshift, without assuming redshift information of the sources. 

In this paper, we further develop the cosmological applications of FRBs-GWs associations with no redshift information. We consider realistic event scenarios for the current and future ground-based detectors, different models of the probability density function (PDF) for the FRBs dispersion measure \citep{Macquart_relation_2020, FRBcosmo_localised_Zhuge_et_al_2025, Connor_2025NatAs, FRB_PDF_Sharma2025}, and employ a  general Bayesian framework to constrain both FRB host properties and an extended set of cosmological parameters, including the DE equation of state. We first take the redshift distribution, simulate GW data for LIGO-Virgo (LV) and CE, and consider the latest PDF for the FRB's dispersion measure to simulate FRB data. We then employ a general Bayesian analysis without relying on redshift or any approximations to the redshift. This represents the most rigorous and general treatment of the subject to date.

The paper is structured as follows: In Section \ref{sec:theory} we review the basics of GWs and FRB cosmology, in Section \ref{sec:simulated_data} we describe the construction of our simulated dataset, in Section \ref{sec:bayesian_analysis} we develop our statistical methodology, in Section \ref{sec:results} we present the cosmological constraints and compare the impact of different detectors and FRB PDF modeling, and in Section \ref{sec:conclusions} we discuss our results and conclusions.

\section{Cosmology measurement from GWs and FRB} \label{sec:theory}

\subsection{Luminosity distances from Gravitational waves} \label{sec:dl_from_GWs}

GWs signals provide a direct measurement of the source's luminosity distance $D_L$, without the need of any external information or a ``distance ladder'' \citep{Maggiore_GWsBook_Vol1}. The GWs amplitude $h$ is to leading order given by
\begin{equation}
    h \sim \frac{4}{D_L} \left( \frac{G \mathcal{M}_c}{c^2} \right)^{5/3} \left( \frac{\pi f_{\rm gw}}{c} \right)^{2/3},
\end{equation}
where $\mathcal{M}_c = (1+z)M_c = (1+z)(m_1 m_2)^{3/5}/(m_1+m_2)^{1/5}$ is the redshifted chirp mass, with $z$ the redshift of the source. $m_1, m_2$ are the masses of the individual objects in their source frame. $f_{gw}$ is the GWs frequency in the rest frame of the detectors on Earth and $D_L$ is the luminosity distance. The latter depends on cosmology through
\begin{equation}\label{eq:luminosity_distance}
    D_L=(1+z)\frac{c}{H_0}\int_0^z \frac{\dd{z'}}{E(z')}.
\end{equation}
In a $\Lambda$CDM cosmology, $E(z) = \sqrt{\Omega_m(1+z)^3+\Omega_\Lambda}$, where we have neglected any contributions from relativistic species and curvature. $\Omega_m$ and $\Omega_\Lambda$ are the dimensionless cosmological densities of matter and dark energy, treated as a cosmological constant, at $z=0$.

For the more general approach we consider here, we allow a more dynamical dark energy, instead of the cosmological constant with $w=-1$ in the standard $p=w \rho c^2$ Equation of State (EoS) parameterization \citep{Hobson_et_al_GRbook}. Hence, for a general DE EoS $w$, the $E(z)$ in a $w$CDM cosmology, with a flat universe, will change as \citep{Linder_DE_eos_2003, Frieman_Turner_Huterer_DE_2008}:

\begin{equation}\label{eq:Ez_wCDM}
    E(z, \Omega_m, w) = \sqrt{\Omega_m (1+z)^3 + (1-\Omega_m)(1+z)^{3(1+w)}}.    
\end{equation}

In an ideal scenario, a precise distance $D_L$ determination and the identification, i.e. measuring the redshift $z$, of the host would only leave the cosmological parameters as unknowns. In reality, the redshift estimation requires usually the observation of an electromagnetic counterpart, while the luminosity distance inferred by GWs is degenerate with other parameters. One of the main factors that weakens the precision of $D_L$ is the well-known degeneracy between inclination and distance \citep{Mastrogiovanni_Steer_2020_GWcosmo_review}. Inclination refers to the angle between the line of sight from the GWs source to the observer and the total orbital angular momentum vector, i.e. an `edge-on' observation corresponds to inclination $i=90$ degrees, while a `face-on' observation has $i=0$.

In this work we exploit the luminosity distance posteriors from realistic GWs events to obtain constraints on $w$CDM. We discuss the GWs data generation and analysis in Section \ref{sec:simulated_data}, and the impact of inclination in our results in App. \ref{App:viewing_angle}.

\subsection{Dispersion measure in Fast Radio Bursts} \label{sec:dm}

FRBs are short, transient signals within radio frequencies \citep{Petroff_FRB_review_2022, Physics_FRBs_Bing_2023}, first discovered by \cite{Lorimer_burst_2007} in archival data from pulsar surveys. They are received as dispersed pulses, where the level of dispersion is calculated by the arrival time difference between the lowest and highest frequencies of the pulse, $\nu_l$ and $\nu_h$ respectively:

\begin{equation}
    \Delta t = \frac{e^2}{2 \pi m_e c} \frac{1}{\nu_l^2-\nu_h^2} \DM,
\end{equation}
where $e, m_e$ are the charge and mass of the electron, $c$ is the speed of light, and $\DM$ is the dispersion measure of the pulse. The latter depends on the amount of free electrons $n_e(z)$ along the path, and is given by integrating the line-of-sight, proper distance $l$ between FRB source $l=0$, and observer $l=L$:

\begin{equation}
    \DM = \int_0^L \frac{n_e(z)}{1+z} \dd{l}.
\end{equation}

During the propagation of the signal, the FRB pulse passes through various environments, whose dispersion measures we analyse separately. Specifically, we can break the observed $\DM$ of an FRB \citep[e.g.][]{FRB_DM_break_Thornton_et_al_2013, FRB_cosmo_DM_Deng_Zhang_2014, 2019_Prochaska_Zheng} to:

\begin{equation}
    \DM_{\rm obs} = \DM_{\rm MW} + \DM_{\rm diff} + \frac{\DM_{\rm host}}{1+z}.
    \label{eq:FRB_DM}
\end{equation}
The $\DM_{\rm MW}$ quantifies the contributions from the Milky Way (MW), and is further separated to $\DM_{\rm MW, ISM}$ and $\DM_{\rm MW, halo}$. The first term describes the effects of the MW interstellar medium and is obtained by electron density models calibrated by radio pulsar data \citep{2002_NE2001I, 2017_YMW16}. The second term incorporates the effects of the extended MW halo and is estimated to be in the range of $\DM_{\rm MW, halo} \sim (30-80) {\rm pc/cm^3}$ \citep{DM_MW_halo_Dolag_et_al_2015}. 

The $\DM_{\rm diff}$ measures the contributions of diffused electrons along the line-of-sight, drawn from the intergalactic medium (IGM) and intervening cosmic halos. This part has the most relevance for cosmological studies, since a mean diffuse $\DM$ can be derived theoretically, assuming a homogeneous distribution and ionization of baryons, with ionized fraction $\chi(z)$. The result depends on various cosmological parameters \citep{DMdiff_Ioka_2003, DMdiff_Inoue_2004, FRB_cosmo_DM_Deng_Zhang_2014}:

\begin{equation}\label{eq:DM_diff}
    \langle{\DM_{\rm diff}}\rangle 
    (z) = \frac{3cH_0\Omega_b }{8\pi G m_p}\int_0^z \frac{ f_{\rm diff}(z')\chi(z')(1+z')}{E(z')}\dd{z'},
\end{equation}
with
\begin{equation}
    \chi(z) = Y_H X_{\rm e, H}(z) + \frac{1}{2}Y_p X_{\rm e, He}(z).
\end{equation}

The different terms above have the following meaning: $G$ Newton's gravitational constant, $m_p$ the mass of the proton, $\Omega_b$ the dimensionless cosmological density of baryons. $f_{\rm diff}$ introduces the astrophysical effects, quantifying the fraction of diffuse baryons in both the IGM and halos. Unless otherwise stated, we follow observational studies that have constrained $f_{\rm diff} \sim 0.84$ with no evidence of redshift dependence \citep{Li_et_al_fdiff_2020}. $\chi(z)$ represents the fraction of ionized electrons over the baryons in the Universe. Its components $Y_i$ correspond to the mass fraction of hydrogen H or helium He in the Universe, and to the ionization fraction of each element $X_{\rm e, i}$. For low redshifts $(z<3)$, the Universe is ionized, and $\chi(z)$ tends to a constant value $\chi(z) \sim 7/8$ \citep{Fukugita_Hogan_Peebles_1998, FRB_cosmo_DM_Deng_Zhang_2014}. $E(z')$ depends on the exact cosmological model, as shown in Section \ref{sec:dl_from_GWs}. 

Lastly, the $\DM_{\rm host}$ assesses the contributions of the host environment, which involves both the immediate source surroundings, and the host galaxy influence. The $(1+z)$ factor is due to the cosmological time dilation for a source at redshift $z$.

\section{Simulated datasets}\label{sec:simulated_data}

Combining the $D_L$ from GWs and $\DM$ from FRBs, one can better constrain the cosmological parameters, especially for associated events. Since there are no confirmed FRB-GW associations at the moment of writing, we exploit simulated FRB and GWs data. To create our mock dataset, we perform the following steps:

\subsection{Identify potential common sources for FRB and GWs events}\label{sec:FRBs_GWs_sources}

Mergers of compact binary objects — binary black holes (BBH), binary neutron stars (BNS), and neutron star-black hole (NS-BH) systems — are potential joint sources of GWs and FRBs. The BBH scenario is generally not expected to emit electromagnetic radiation; thus, despite speculative models such as charged black holes \citep{BHs_charged_Zhang_2016}, it is an unlikely candidate for sourcing FRBs. Conversely, while the BNS scenario is often considered a favorable theoretical model due to its strong magnetic fields \citep{FRBGW_BNs_2001MNRAS.322..695H, FRBGW_BNs_2012ApJ...755...80P, FRBGW_BNs_2014ApJ...780L..21Z, FRBGW_BNs_2016ApJ...822L...7W, FRBGW_BNs_2019ApJ...886..110M, FRBGW_BNs_2020ApJ...891...72W}, it faces significant observational challenges. The ``dirty'' environment inherent to BNS mergers may impede the propagation of FRBs \citep{BNS_environment_Fong_et_al_2015, BNS_environment_Lu_et_al_2021}. If an FRB successfully escapes, it is likely associated with a relativistic jet (and thus a Gamma-Ray Burst), which typically provides redshift and inclination information \citep{FRBGW_BNs_2014ApJ...780L..21Z,FRB_cosmo_DM_Deng_Zhang_2014, BNS_MMA_followUp_LVK_et_al_2018, Chen_at_al_GWs_angle_2019, BNS_MMA_followUp_Colombo_et_al_2022, Yiying_Wang_GRB_GW_viewing_angle}. 
Since redshift is measured independently in such a scenario, it does not apply to our discussed methodology, so we do not consider BNS mergers as the primary population in this study.

Instead, we focus on NS-BH mergers \citep{FRBGW_BHNS_2015ApJ...814L..20M, Zhang_2019,FRBGW_BHNS_2019ApJ...873L..13D, FRBGW_BHNS_2023ApJ...956L..33M}, specifically ``plunging events'' with a small mass ratio, $q = M_{\rm NS}/M_{\rm BH} \ll 1$. In this scenario, the neutron star is swallowed by the black hole without tidal disruption, ensuring a ``clean environment'' for the FRB to propagate \citep{MMA_FRBS_Bing_2024}. This choice provides a more rigorous test for our framework: as NS-BH events often lack identifiable electromagnetic counterparts beyond the FRB \citep{NSBH_EM_Counterparts_Zhu_et_al_2021, NSBH_merger_rates_Zhu_et_al_2021}, they are affected by significant uncertainties on their redshifts. Furthermore, since our method yields robust cosmological constraints in this conservative NS-BH scenario without redshift, it follows that BNS mergers — which offer richer multi-messenger data — would inherently lead to even more precise results.

\subsection{Redshift distribution}\label{sec:Redshift_distribution}

Having identified our potential sources for FRBs and GWs associations, we now investigate their redshift distribution. The astrophysical population of the latest GWs detections \citep{GWTC4_Catalogue_2025, GWTC4_Population_2025} includes fewer than $10$ NS-BH events, and all of them with estimated redshifts $z \lesssim 0.2$. For this reason, we base our redshift distribution to theoretical modeling of NS-BH events, following \citet{NSBH_merger_rates_Zhu_et_al_2021}. Their redshift distribution is motivated from cosmological star formation history and time-delays for formation and merger of these systems. They consider three merger delay models from the literature: the Gaussian delay model \citep{Rates_Gaussian_Virgilli_et_al_2011}, the log-normal delay model \citep{Rates_LogNormal_PowerLaw_Wanderman_Piran_2015} and the power-law delay model \citep{Rates_LogNormal_PowerLaw_Wanderman_Piran_2015}.

A comparison of the different cases is provided in Appendix \ref{App:mock_frb_gws_catalogues}. Given the large uncertainties in the underlying models and their similar cosmological constraints in idealized scenarios (see Section \ref{App:Ideal_observations}), we adopt the power-law delay model \citep{Rates_LogNormal_PowerLaw_Wanderman_Piran_2015} as our fiducial choice. The corresponding redshift distribution, further detailed in \citet{NSBH_merger_rates_Zhu_et_al_2021}, provides a robust and representative description of the current NS-BH redshift distribution:
\begin{equation}\label{eq:power_law_redshift}
f_{\rm PL}(z) = \Bigg[ (1+z)^{1.895 \eta} +\left( \frac{1+z}{5.722}\right)^{-3.759 \eta} + \left( \frac{1+z}{11.55}\right)^{-0.7426 \eta} \Bigg]^{1/\eta},
\end{equation}
where $\eta=-8.161$.

For the redshift range of our sources, we assume $z \leq 3$, for two main reasons: (1) We want to be in the range where H and He are fully ionized, in order for the approximation $\chi(z) \sim 7/8$ to hold, and (2) we want to be within the ranges of current and future GWs detectors (for more details, see Section \ref{sec:GW_gen}).

The NS-BH  rates are very uncertain for this redshift range both for current and future detectors. Current observations \citep{GWTC4_Population_2025} give a merger rate $\mathcal{R}_{\rm NSBH} = 9.1-84$ Gpc$^{-3}$ yr$^{-1}$, while estimates for future detectors \citep[e.g.][]{NSBH_merger_rates_CE_MG_Chen_et_al_2024} predict as much as $10^4$ events for 1 year, and about $50$ for very small inclinations ($\leq 0.1$ rad). The later estimate is about the same order of magnitude of a simple extrapolation of the current merger rates to $z=3$, i.e. $D_L \simeq 26$ Gpc, $\mathcal{R}_{\rm NSBH} = 6 \cdot10^5$ yr$^{-1}$. Finally, the CE is expected to detect almost all $1.4 \ M_\odot$ BNS till $z \sim 10$, and as a result will be able to cover the heavier NSBH systems \citep{CE_sensitivity_2021}. Hence, we conclude that the number of events we consider in this paper, $N_{\rm events} = 50$, is a realistic value for current capabilities, and a conservative value for future detectors.

\subsection{$D_L$ generation}\label{sec:GW_gen}

On the GW side, in order to get realistic luminosity distance uncertainties, we inject NS-BH mergers in a current (LIGO-Virgo) \citep{BNS_MMA_followUp_LVK_et_al_2018} and future (Cosmic Explorer) \citep{Sensitivity_future_detectors_2017, CE_sensitivity_2021, CE_sensitivity_2022} gravitational wave detector network\footnote{The detector noise curves used in the simulations are provided by \texttt{Bilby} and can be found for Virgo [\href{https://dcc.ligo.org/LIGO-P1200087-v42/public}{here}], for LIGO [\href{https://dcc.ligo.org/LIGO-T2000012/public}{here}] and for CE [\href{https://dcc.ligo.org/LIGO-P1600143/public}{here}].}. We perform our simulations using \texttt{Bilby} \citep{Bilby_paper_2019, Bilby_cbc_2020}. We inject a NS and BH with source frame masses $M_{\rm NS} = 1.4\ M_\odot$ and $M_{\rm BH} = 14\ M_\odot$, in order to have a mass ratio of $q=1/10$, consistent with the ``plunging'' type events (Section \ref{sec:FRBs_GWs_sources}). The sources are placed at a series of redshifts $z\ \epsilon\ \{0.05, 0.1, 0.2, 0.5, 0.75, 1, 1.5, 2, 2.5, 3\}$, and are assumed to be well-localized on their angular position on the sky\footnote{This is a valid assumption for the FRB-GWs associations that we consider here. Since FRB localization is much better than GWs one, the former could be used as the injected location for a GWs parameter estimation analysis.}. Assuming a detection threshold of ${\rm SNR}=8$, the LV network can reach till $z \leq 0.2$, while the CE detector is able to confidently detect these events for all the $z$-set. To simplify the analysis, we have neglected any spins on the compact objects. We have set the injected viewing angle to $\theta_{\rm JN} = 0.4$ rad, which in the case of zero spins is equivalent to the inclination angle described in Section \ref{sec:dl_from_GWs}. We discuss the possible effects of the viewing angle choice to the final results in App.~\ref{App:viewing_angle}.

Our results are: $3$ NS-BH simulations for the LV network ($z=0.05, 0.1, 0.2$) and $8$ NS-BH simulations for the CE detector ($z=0.05, 0.1, 0.2, 0.5, 0.75, 1, 1.5, 2, 2.5, 3$). For each event, we have posterior samples for each physical parameter, but crucially for this work, for the luminosity distance of the events. As a sanity check, we confirmed that the input values are recovered in all cases.

\begin{figure}
\centering
\includegraphics[width=\textwidth]{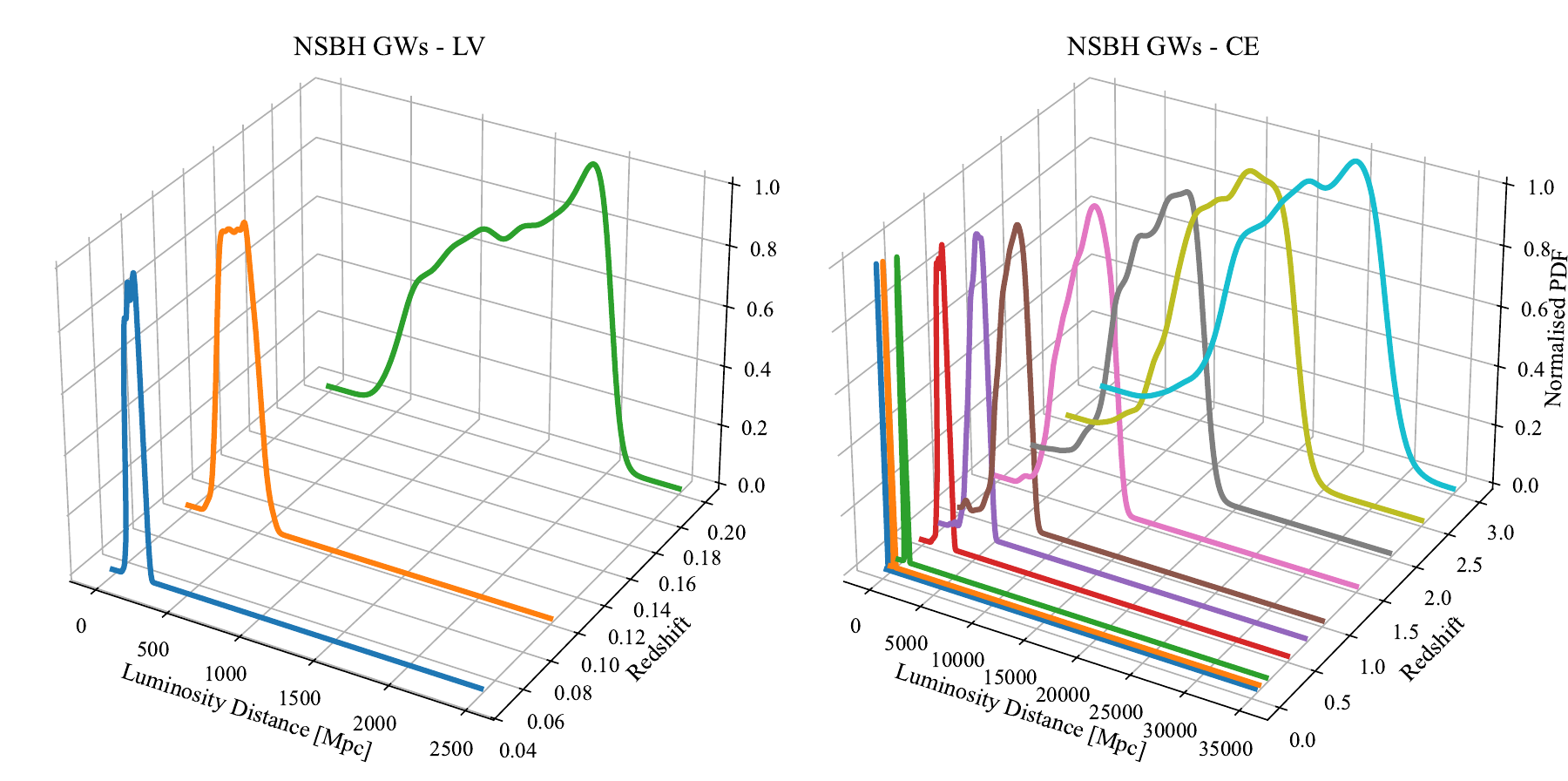}
\caption{Luminosity distance likelihoods for simulated NSBH GWs events, for a LV network (left) and for a CE detector (right). All of the distributions were normalised so that they have the same maximum height. Their widths are used to estimate how the $D_L$ uncertainty scales with redshift.}
\label{fig:DL_posteriors_LV_CE}
\end{figure}

We exploit the realistic $D_L$ posteriors to estimate how the relative error on luminosity distance $E_{D_L} = \Delta_{D_L}/D_L$ changes with redshift. As a first step, we correct the effect of the $D_L^2$ prior we used in the parameter inference and collect the $D_L$ likelihoods for each redshift (see Figure \ref{fig:DL_posteriors_LV_CE}). We then calculate a measure of width for each distribution: we have compared a standard deviation estimate ($\Delta_{D_L}=\sigma_{D_L}$) and an interquartile ($\Delta_{D_L}$ = IQR = $Q3-Q1$) estimate which give equivalent results for our desired level of complexity. Finally, we interpolate the relative error in the redshift range $z \ \epsilon \ [0, 3]$, where we generate our events (Section \ref{sec:generating_mock_events}). The interpolated errors are also used in the cosmological inference (Section \ref{sec:bayesian_analysis}).

\subsection{$\DM$ generation}\label{sec:DM_gen}

FRB sources, their environments and the propagation of FRB signals include a number of uncertainties. As mentioned in Section \ref{sec:dm}, the observed dispersion measure depends on a number of contributions. In real observations, the MW ISM component is the easiest to be subtracted by exploiting the NE2001 \citep{2002_NE2001I, 2003_NE2001II} or YMW16 \citep{2017_YMW16} models. Although the MW halo component remains poorly constrained, its uncertainty and spatial variations are implicitly absorbed into the distribution of $\DM_{\rm diff}$, whose empirically calibrated variance is broad enough to account for these residual Galactic fluctuations. Thus for the simulated $\DM$s, we assume that the different MW components have been subtracted. The remaining terms are the extragalactic $\DM$ defined as $\DM_{\rm ext}=\DM_{\rm diff}+\DM_{\rm host}/(1+z)$. In this project, we consider two scenarios for the FRBs dispersion measure:

In the first scenario, we assume that the contributions of the host have been identified and subtracted. As a result, we consider only the ${\DM_{\rm diff}}$ term. This dataset also acts as a test-case for smaller uncertainty, i.e. more precise measurements. For the $\DM_{\rm diff}$ PDF, we use \citep{McQuinn_2014, 2019_Prochaska_Zheng, Macquart_relation_2020}:
\begin{equation}
    p_{\rm \Delta, \rm Mq}(\Delta)={\cal A} \Delta^{-\beta}\exp\left[-\frac{(\Delta^{-\alpha}-C_0)^2}{2\alpha^2\sigma_{\rm diff}^2}\right],
    \label{eq:p_Delta_Mq}
\end{equation}
where $\Delta={\DM_{\rm diff}}/\langle{\DM_{\rm diff}}\rangle$ is the dimensionless fraction between the measured and theoretical diffuse $\DM$ terms; ${\cal A}$ is the normalization factor; $C_0$ is a factor imposed by the condition that the mean value of $\Delta$ equals $1$, which holds by definition; $\alpha$ and $\beta$ are parameters that describe the gas profile in cosmic halos and we adopt $\alpha=\beta=3$ \citep{Macquart_relation_2020}. Finally $\sigma_{\rm diff}(S, z)$ is related to the scatter of $\Delta$ and we take the more realistic form derived in our previous paper \citep{FRBcosmo_localised_Zhuge_et_al_2025}. This PDF only works for redshifts $z>0.2$ \citep{FRBcosmo_localised_Zhuge_et_al_2025}, thus it can only be used for Cosmic Explorer.

On the other hand, some studies \citep{FRB_PDF_Konietzka2025, Connor_2025NatAs, FRB_PDF_Sharma2025} suggest a log-normal distribution for $\DM_{\rm diff}$. The PDF is:
\begin{equation}
    p_{\rm \Delta,ln}(\Delta|\mu_{\rm ln}, \sigma_{ln})=\frac{1}{\sqrt{2\pi}\sigma_{ln}\Delta}\exp[-\frac{(\ln{\Delta-\mu_{\rm ln}})^2}{2\sigma_{ln}^2}],
    \label{eq:p_Delta_ln}
\end{equation}
where the redshift-dependent parameters $\mu_{\rm ln}$ and $\sigma_{\rm ln}$ are determined by a fit to the results from \citet{FRB_PDF_Sharma2025} and ensuring the distribution has a mean of unity. We specifically adopt the fit corresponding to the largest $\sigma_{\rm ln}$ scenario from the \verb|Astrid| \citep{Astrid} simulation to account for maximal uncertainties. Eq. (\ref{eq:p_Delta_ln}) holds for all redshifts, which allows us a comparison between the two detectors (LV vs CE) for $z<0.2$ events, as well as a comparison of PDF modeling (Eq. (\ref{eq:p_Delta_Mq}) vs Eq. (\ref{eq:p_Delta_ln})) for $z>0.2$ events observed by CE.

In the second scenario, we consider a more realistic case, where the properties of the host are uncertain. This adds some additional astrophysical parameters in our inference, which currently are difficult to be modeled and subtracted. Moreover, during the merger process, the immediate environment may change dramatically, which makes the prior modeling of these terms even harder. For the $\DM_{\rm host}$, we take a log-normal distribution \citep{Macquart_relation_2020}:
\begin{equation}
    p_{\rm host}({\DM_{\rm host}|\mu, \sigma_{\rm host}})=\frac{1}{\sqrt{2\pi}\sigma_{\rm host}{\DM_{\rm host}}}\exp[-\frac{(\ln{\DM_{\rm host}-\mu})^2}{2\sigma_{\rm host}^2}],
    \label{eq:p_host}
\end{equation}
which has an asymmetric tail, allowing potentially large values of $\DM_{\rm host}$ due to dense local conditions. The median value for this function is $\exp(\mu)$. The variance of this function is $[\exp(\sigma_{\rm host}^2)-1]\exp(2\mu+\sigma_{\rm host}^2)$. When we need to assume any fiducial values for $\mu, \sigma_{\rm host}$, we choose $\exp(\mu) = 182.937$ pc/cm$^3$ and $\sigma_{\rm host} = 0.605$ based on \citep{FRBcosmo_localised_Zhuge_et_al_2025}. Note that the variable $\DM_{\rm host}$ here should be the value in the source frame, because in Eq. (\ref{eq:FRB_DM}) we divide $\DM_{\rm host}$ by $(1+z)$.

In summary, we simulate $\DM$ data for two cases: First, we consider the idealized scenario focusing only on $\DM_{\rm diff}$, assuming that all other contributions have been perfectly removed. This dataset will have much smaller uncertainties and should lead to more precise constraints. This case corresponds to a future setting, where we can model and subtract the $\DM_{\rm host}$ term from the host galaxy and immediate environment. Second, we assume a combination of two terms for the extragalactic $\DM$, i.e. $\DM_{\rm ext}=\DM_{\rm diff}+\DM_{\rm host}/(1+z)$. In this case, we model the combined PDF of the extragalactic DM ($\DM_{\rm ext}$) \citep{FRBcosmo_localised_Zhuge_et_al_2025} as
\begin{equation}\label{eq:convolution_pDM_FRB}
  p_{\rm ext}(\DM_{\rm ext}) = \frac{1}
  {\langle{\DM_{\rm diff}}\rangle(z)}
  \int_0^{\DM_{\rm ext}(1+z)} p_{\rm host}(\DM_{\rm host}) p_\Delta \left( \frac{\DM_{\rm ext}-\DM_{\rm host, z}}
  {\langle{\DM_{\rm diff}}\rangle(z)}
  \right) \dd{\DM_{\rm host}},
\end{equation}
where $\DM_{\rm host,z}=\DM_{\rm host}/(1+z)$.

\subsection{Generating mock FRB and GWs sources}\label{sec:generating_mock_events} 

Taking all of the above into consideration, the generation of our dataset of FRB and GWs sources follows the subsequent steps:

\begin{enumerate}
    \item A redshift is drawn randomly between $0<z<0.2$, or $0.2\leq z \leq 3$, following the distribution described in Section \ref{sec:Redshift_distribution}.
    
    \item For each redshift, we calculate the theoretical $D_L^{\rm th}$, $\DM_{\rm diff}^{\rm th}$ and $\DM_{\rm ext}^{\rm th}$. We then generate the observed $D_L^{\rm obs}$, $\DM_{\rm diff}^{\rm obs}$ and $\DM_{\rm ext}^{\rm obs}$ based on their PDFs: $\DM_{\rm diff}^{\rm obs} \sim p_{\Delta}(\DM_{\rm diff}^{\rm th}, \sigma_{\rm diff})$, $\DM_{\rm ext}^{\rm obs} \sim p_{\rm ext}(\DM_{\rm ext}^{\rm th}, \sigma_{\rm diff}, \sigma_{\rm host})$, while, for simplicity, we assume for GWs' luminosity distance a Gaussian, i.e. $D_{L}^{\rm obs} \sim \mathcal{N}(D_L^{\rm th}, \sigma_{D_L})$.
    
    \item We repeat this procedure $N_{\rm events}=50$ times, to get a dataset of $N_{\rm events}$ pairs ($\DM_{\rm diff}^{\rm obs}, D_{L}^{\rm obs}$) and ($\DM_{\rm ext}^{\rm obs}, D_{L}^{\rm obs}$).
    
    \item We use these pairs to infer cosmological constraints. 
\end{enumerate}

\begin{figure}
    \centering
    \includegraphics[width=\linewidth]{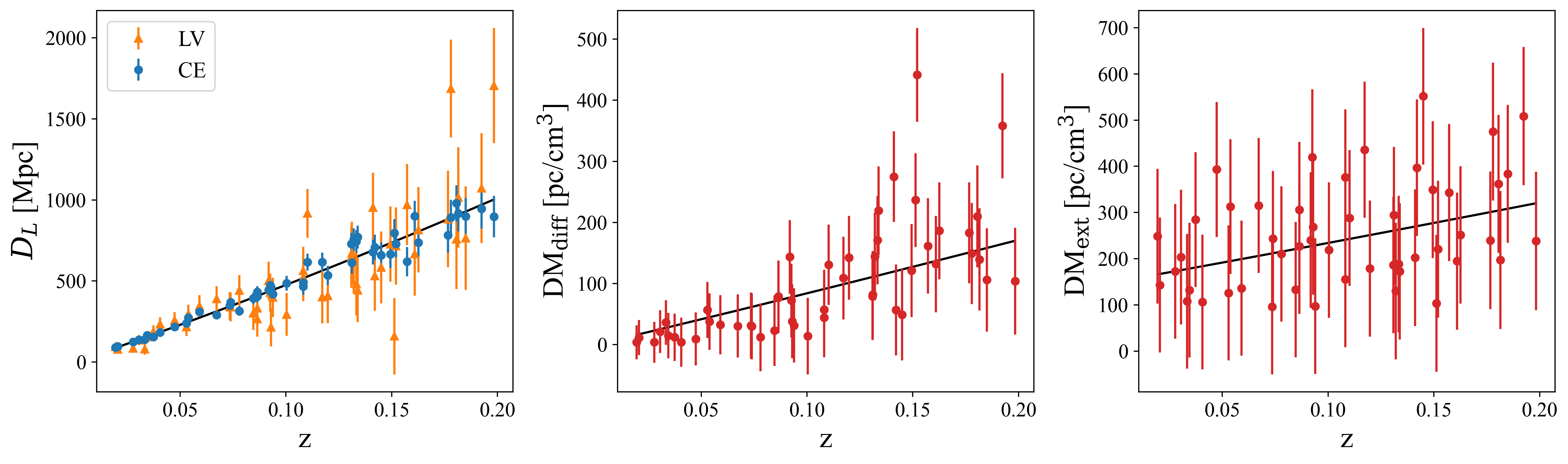}
    \caption{Simulated GWs and FRBs events as a function of redshift ($z<0.2$). (\emph{Left}) GWs luminosity distance, observed by LV or CE. (\emph{Center}) FRBs diffuse dispersion measure, $\DM_{\rm diff}$. In this case, we assume that one is able to remove any other contributions to the $\DM$ and keep only the cosmological ones. (\emph{Right}) FRBs extra galactic dispersion measure, $\DM_{\rm ext}$. The latter involves cosmological and host galaxy contributions. For fiducial testing, we generate $N=50$ events.}
    \label{fig:events_DL_DM_redshift_distribution_02}
\end{figure}

\begin{figure}
    \centering
    \includegraphics[width=\linewidth]{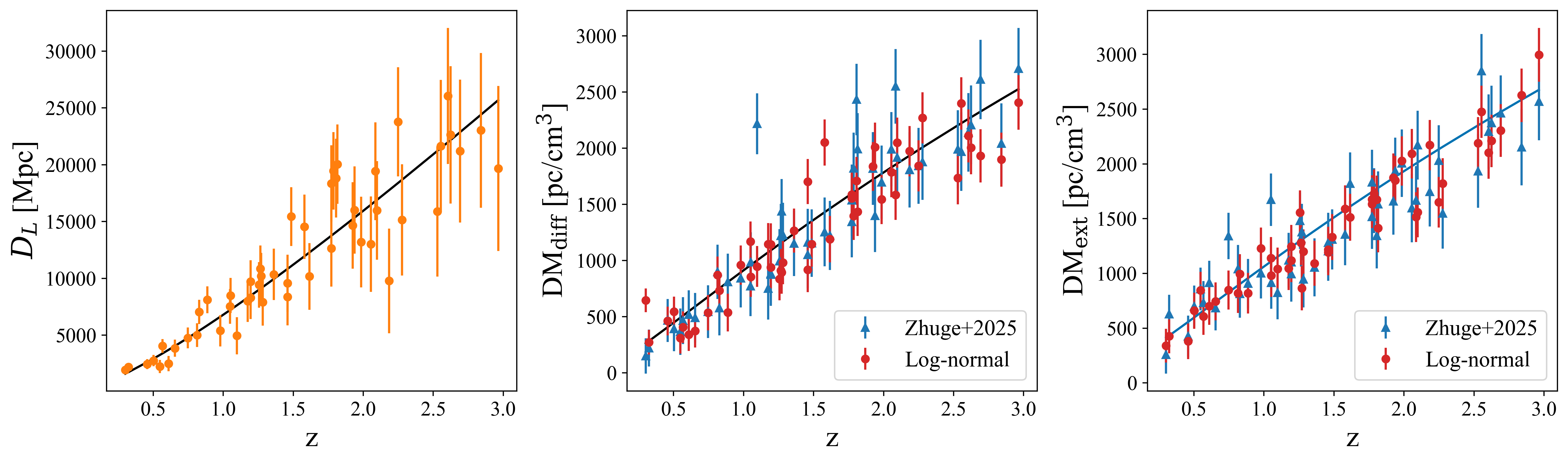}
    \caption{Simulated GWs and FRBs events as a function of redshift ($0.2<z<3$). (\emph{Left}) GWs luminosity distance, observed by or CE. (\emph{Center}) FRBs diffuse dispersion measure, $\DM_{\rm diff}$. In this case, we assume that one is able to remove any other contributions to the $\DM$ and keep only the cosmological ones. (\emph{Right}) FRBs extra galactic dispersion measure, $\DM_{\rm ext}$. The latter involves cosmological and host galaxy contributions. For fiducial testing, we generate $N=50$ events.}
    \label{fig:events_DL_DM_redshift_distribution_3}
\end{figure}

For the input cosmological parameters we follow \verb|Planck18| \citep{Planck_Cosmo_param_2018}, i.e. ($H_0, \Omega_m, w$) = ($67.66$ km/s/Mpc, $0.30966, -1$). We show the simulated $D_L,\ \DM_{\rm diff}$ and $\DM_{\rm ext}$ and their uncertainties as a function of redshift in Figure \ref{fig:events_DL_DM_redshift_distribution_02} ($z<0.2$) and Figure \ref{fig:events_DL_DM_redshift_distribution_3} ($0.2<z<3$) for our mock GWs-FRB events.

\section{Bayesian analysis}\label{sec:bayesian_analysis}

We develop a Bayesian analysis to get cosmological constraints and astrophysical constraints by combining GWs and FRB detections, without knowledge of any redshift information. We start by considering the case of known FRB host properties in Section \ref{sec:BA_cosmo_known_FRB_host}, while we expand to the full model in Section \ref{sec:BA_cosmo_unknown_FRB_host}. As a comparison and test for Gaussian distributions, we also do a $\chi^2$ analysis in App.~\ref{App:Ideal_observations}, while in App.~\ref{App:Low_z_limit} we provide a simple scaling relation for $H_0$ as a function of $D_L$ and $\DM_{\rm diff}$ that holds only at low redshifts.

\subsection{Cosmological constraints - known FRB host properties}\label{sec:BA_cosmo_known_FRB_host}

In an ideal observation, when we know, or can estimate the FRB host properties, we can disentangle the part of the $\DM$ that carries cosmological information $\DM_{\rm diff}$. We start with a single event, and since we do not know the redshifts, we marginalize over them:
\begin{align}
    p( \mathbf{H} | \mathbf{D}, M) &= \int   p( \mathbf{H}, z | \mathbf{D}, M) dz \propto \int p(\mathbf{D} | \mathbf{H}, z, M) p( \mathbf{H}, z | M) \dd{z} \nonumber \\
    & = \int p(\mathbf{D} | \mathbf{H}, z, M) \pi( z | \mathbf{H},  M) \pi(\mathbf{H} |M) \dd{z} \nonumber \\
    & = \pi(\mathbf{H} |M) \int p(D_L^{\rm obs}, \DM_{\rm diff}^{\rm obs} | \mathbf{H}, z, M) \pi( z | \mathbf{H}, M)  \dd{z} \nonumber \\
    & = \pi(\mathbf{H} |M) \int p(D_L^{\rm obs} | \mathbf{H}, z, M) p(\DM_{\rm diff}^{\rm obs} | \mathbf{H}, z, M) \pi( z | \mathbf{H}, M)  \dd{z}.
\end{align}

Here we denote the cosmological parameters vector as $\mathbf{H}= \{H_0, \Omega_m, w\}$, the data vector $\mathbf{D}= \{ D_L^{\rm obs}, \DM_{\rm diff}^{\rm obs} \}$, while $M$ denotes the different model assumptions, i.e. the form of the PDF chosen for $D_L$ and $\DM_{\rm diff}$, and the redshift prior $\pi(z)$. We will describe our choices explicitly below. In the derivation above, we use Bayes' theorem in the first line. In the second line, the joint prior of the redshift and cosmological parameters is decomposed into their respective priors using the chain rule. Since $\pi(\mathbf{H}|M)$ is independent of the integration variable $z$, it can be treated as a constant and factored out of the integral. Finally, in the last step, we assume that the GWs $D_L$ and FRB $\DM$ are independent observations, i.e. observed by different facilities.

For multiple, independent events, the data vector becomes $\mathbf{D}^i= \{ D_L^{\rm obs, i}, \DM_{\rm diff}^{\rm obs, i} \}$, where $i$ denotes the single event. For $N$ events, this leads to 
\begin{equation}
    p(\mathbf{D} | \mathbf{H}, \mathbf{z}, M) = \prod_i^N p(\mathbf{D}^i | \mathbf{H}, z_i, M) = \prod_i^N p(D_L^{\rm obs, i}, \DM_{\rm diff}^{\rm obs, i} | \mathbf{H}, z_i, M).
\end{equation}

Note here that the redshift $z_i$ is a specific parameter for each event, not a global one, since different events are expected to have independent distances. Below we marginalize the redshift uncertainty for each event separately creating a `marginalized likelihood' for each observation \citep{Hogg_et_al_2010_marg, Loredo_Hendry_2019_marg, Thrane_Talbot_2019_marg}. Therefore, the final posterior is given by
\begin{align}\label{eq:BA_posterior_multiple_events_known_FRB_host}
    p( \mathbf{H} | \mathbf{D}, M) &\propto \pi(\mathbf{H} |M) \prod_i^N \int p(D_L^{\rm obs, i}, \DM_{\rm diff}^{\rm obs, i} | \mathbf{H}, z_i, M) \pi( z_i | \mathbf{H}, M) \dd{z_i} \nonumber \\
    &= \pi(\mathbf{H} |M) \prod_i^N \int p(D_L^{\rm obs, i} | \mathbf{H}, z_i, M) p(\DM_{\rm diff}^{\rm obs, i} | \mathbf{H}, z_i, M) \pi( z_i | M) \dd{z_i}.
\end{align}

Eq. (\ref{eq:BA_posterior_multiple_events_known_FRB_host}) allows the determination of cosmological parameters starting from the observed ``distance'' properties of the FRB-GWs pair. The term $\pi(\mathbf{H} |M)$ describes the priors on the cosmological parameters, which are chosen to be uniform in this work and within ranges described in Table \ref{tab:param_priors}. The $M$ in the PDFs signifies that specific model choices need to be made. For $p(\DM_{\rm diff}^{\rm obs, i} | \mathbf{H}, z, M)$ we use Eq. \ref{eq:p_Delta_Mq} and Eq. \ref{eq:p_Delta_ln}. Since this is the PDF for the variable $\Delta$, a factor of $1/\DM_{\rm diff}^{\rm th}$ is required to convert it into the probability distribution for DM. For $p(D_L^{\rm obs, i} | \mathbf{H}, z, M)$ we use a Gaussian $\mathcal{N}(D_L^{\rm obs, i} - D_L(z, \mathbf{H}), \Delta_{D_L}(z))$, where the luminosity distance $D_L(z, \mathbf{H})$ depends on cosmology (section \ref{sec:dl_from_GWs}) and $\Delta_{D_L}(z)$ follows realistic uncertainties from simulated GWs events (section \ref{sec:GW_gen}). Finally the redshift prior $\pi(z, \mathbf{H}|M)$ follows the redshift distribution described in section \ref{sec:Redshift_distribution}. A full analysis could leave any cosmological (or astrophysical) dependence in $\pi(z, \mathbf{H}|M)$ and allow it to be constrained in the inference. For simplicity, and since the impact should be small (see App.~\ref{App:mock_frb_gws_catalogues}), we choose a fixed redshift distribution, i.e. $\pi(z|M)$, with $M$ the power-law delay model, Eq. (\ref{eq:power_law_redshift}).

\subsection{Cosmological and host constraints - unknown FRB host properties}\label{sec:BA_cosmo_unknown_FRB_host}

In the case of ($D_L, \DM_{\rm ext}$) where the host galaxy's DM contribution is unknown, the set of parameters must be expanded to include host-related variables. Assuming the host parameter vector is $\mathbf{x}^{\rm host}=(\mu, \sigma_{\rm host})$, the posterior for a single event becomes:

\begin{align}
    p( \mathbf{H},\mathbf{x}^{\rm host} | \mathbf{D}, M) 
    &= \int   p( \mathbf{H}, \mathbf{x}^{\rm host}, z | \mathbf{D}, M) dz 
    \propto \int p(\mathbf{D} | \mathbf{H}, \mathbf{x}^{\rm host}, z, M) p( \mathbf{H}, \mathbf{x}^{\rm host}, z | M) \dd{z} \nonumber \\
    & = \int p(\mathbf{D} | \mathbf{H}, \mathbf{x}^{\rm host}, z, M) \pi( z | \mathbf{H},M) \pi(\mathbf{H},\mathbf{x}^{\rm host} |M) \dd{z} \nonumber \\
    & = \pi(\mathbf{H}, \mathbf{x}^{\rm host} |M) \int p(D_L^{\rm obs}, \DM_{\rm ext}^{\rm obs} | \mathbf{H}, \mathbf{x}^{\rm host}, z, M) \pi( z |\mathbf{H},M)  \dd{z} \nonumber \\
    & = \pi(\mathbf{H}|M)\pi(\mathbf{x}^{\rm host} |M) \int p(D_L^{\rm obs} | \mathbf{H},z, M) p(\DM_{\rm ext}^{\rm obs} | \mathbf{H}, \mathbf{x}^{\rm host}, z, M) \pi( z |M)  \dd{z}.
\end{align}

In the second line we have assumed that the FRB host properties do not change with redshift, while in the final line we explicitly denote that the GWs observations do not depend on the host environment of the FRB. 

The final posterior for multiple events becomes
\begin{equation}\label{eq:BA_posterior_multiple_events_known_cosmo}
    p( \mathbf{H},\mathbf{x}^{\rm host} | \mathbf{D}, M) 
    \propto \pi(\mathbf{H}|M)\pi(\mathbf{x}^{\rm host} |M)  \prod_i^N \int p(D_L^{\rm obs, i} | z_i, M) p(\DM_{\rm ext}^{\rm obs, i} | \mathbf{x}^{\rm host}, z_i, M) \pi( z_i | M) \dd{z_i},
\end{equation}

with the prior choices shown in Table \ref{tab:param_priors}.

\begin{table}[ht]
    \centering
    \begin{tabular}{|c|c|c|}
    \hline
        Parameters & Range & Units\\
        \hline
        \hline
        $H_0$ & $\mathcal{U}(40, 100)$ & km/s/Mpc\\
        $\Omega_m$ & $\mathcal{U}(0, 1)$ & --- \\
        $w$ & $\mathcal{U}(-2, 0)$ & --- \\
        $\exp(\mu)$ & $\mathcal{U}(10, 400)$ & pc/cm$^3$ \\
        $\sigma_{\rm host}$ & $\mathcal{U}(0.2, 1.4)$ & --- \\
    \hline
    \end{tabular}
    \caption{Cosmological and astrophysical parameters priors' distributions and range. $\mathcal{U}$ denotes a uniform distribution.}
    \label{tab:param_priors}
\end{table}

\section{Results and discussion}\label{sec:results}

In this section, we present the results of our Bayesian inference performed via Markov Chain Monte Carlo (MCMC) simulations. We first evaluate the low-redshift data, focusing on a comparative analysis of the constraining power between the LV and CE detector networks. Subsequently, we extend our analysis to the high-redshift case to investigate the systematic impacts of different diffuse electron PDF models on the resulting cosmological and astrophysical constraints.

\subsection{Detector comparison $(z<0.2)$}

To isolate the impact of detector sensitivity on cosmological constraints, we perform a comparison by restricting both LIGO-Virgo (LV) and Cosmic Explorer (CE) to a low-redshift sample ($z < 0.2$). While CE's reach extends far beyond this range, this setting allows us to directly evaluate how the superior precision in luminosity distance ($D_L$) measurement translates into parameter estimation.

\begin{figure}
\centering
\includegraphics[width=1.0\textwidth]{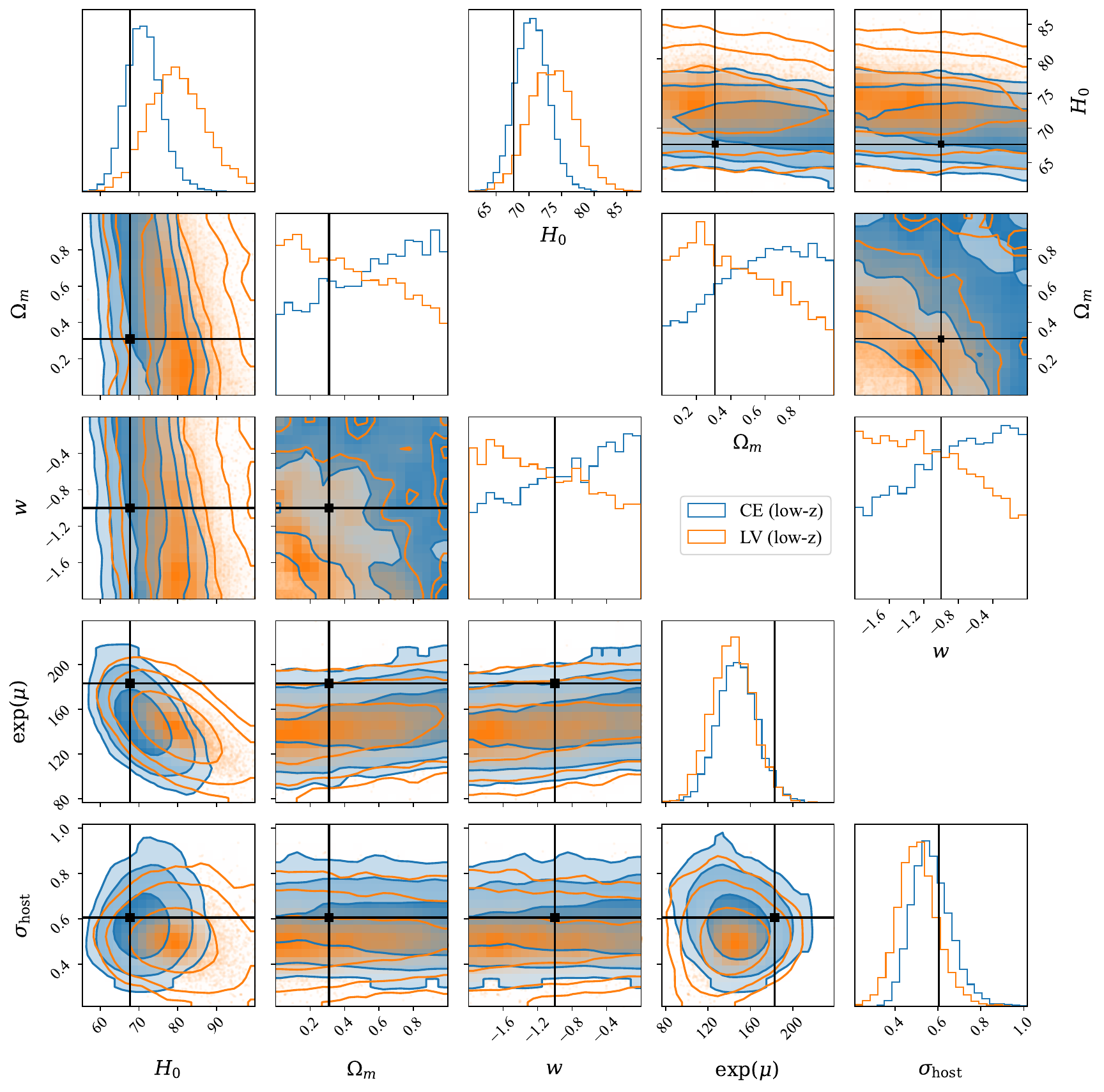}
\caption{Joint constraints on cosmological and host galaxy parameters for low-redshifts ($z<0.2$). The lower-left panels display the results obtained using the $(D_L, \DM_{\rm ext})$ dataset, while the upper-right panels show the constraints from the $(D_L, \DM_{\rm diff})$ dataset. The blue contours represent the constraints from the CE network, while the orange contours correspond to the LV network. Solid black lines indicate the input values from \texttt{Planck18} \citep{Planck_Cosmo_param_2018}.}
\label{fig:MCMC_02}
\end{figure}

In the ideal case where the host galaxy DM is perfectly removed (upper-right panels of Figure \ref{fig:MCMC_02}), the disparity between the $H_0$ constraints of the two detector networks is striking. CE yields a tight, symmetric posterior, whereas the LV result is significantly broader with larger uncertainties. This discrepancy is a direct consequence of the higher SNR and lower $D_L$ error in CE. However, both detectors fail to constrain $\Omega_m$ and $w$ at these low redshifts, as the cosmological volume is too small to break the degeneracies between these parameters. These results suggest that for current-generation detectors (LV), even ``clean'' DM data is insufficient for precision cosmology at low $z$ due to the large distance uncertainties.

When considering the more realistic extragalactic DM, which includes host galaxy contributions, the results in the lower-left panels of Figure \ref{fig:MCMC_02} follow a similar trend. CE maintains a sharper constraint on $H_0$ compared to LV. Notably, the nuisance parameters $\exp(\mu)$ and $\sigma_{\rm host}$ are well-constrained and show nearly identical distributions for both detectors. This indicates that the primary limiting factor for cosmological inference at low redshift is the $D_L$ measurement precision rather than the uncertainties in host galaxy modeling. Consequently, while CE can achieve higher-precision $H_0$ measurements in the local universe, LV requires a significantly larger event sample to compensate for its higher observational noise.

\begin{table}[ht]
    \centering
    \caption{Summary of MCMC constraints for low-redshift events ($ z < 0.2$)}
    \label{tab:MCMC_Lowz}
    \begin{tabular}{lcccccc}
        \toprule
        Data & Detector & $H_0$ [$\text{km/s/Mpc}$] & $\Omega_m$ & $w$ & $\exp(\mu)$ [$\text{pc/cm}^3$] & $\sigma_{\rm host}$ \\
        \midrule
        $D_L-\DM_{\rm diff}$ & LV & $\LowDiffLVKH$ & $\LowDiffLVKOmega$ & $\LowDiffLVKw$ & --- & --- \\
        $D_L-\DM_{\rm diff}$ & CE & $\LowDiffCEH$ & $\LowDiffCEOmega$ & $\LowDiffCEw$ & --- & --- \\
        \addlinespace
        $D_L-\DM_{\rm ext}$ & LV & $\LowExtLVKH$ & $\LowExtLVKOmega$ & $\LowExtLVKw$ & $\LowExtLVKmu$ & $\LowExtLVKsigma$ \\
        $D_L-\DM_{\rm ext}$ & CE & $\LowExtCEH$ & $\LowExtCEOmega$ & $\LowExtCEw$ & $\LowExtCEmu$ & $\LowExtCEsigma$ \\
        \bottomrule
    \end{tabular}
\end{table}

In this low-redshift regime, we focus on the comparison between current and next-generation detector sensitivities. The quantitative results for all considered scenarios, including both $D_L-\DM_{\rm diff}$ and $D_L-\DM_{\rm ext}$ relations for the LIGO-Virgo (LV) and Cosmic Explorer (CE) networks, are summarized in Table \ref{tab:MCMC_Lowz}.

\subsection{Future CE detector $(0.2<z<3)$}

While the previous subsection demonstrated that only the Cosmic Explorer (CE) can provide meaningful constraints within the local universe ($z < 0.2$), the operational range of both CE and FRB observations extends significantly further, reaching up to $z \approx 3.0$. In this subsection, we investigate the cosmological potential of GW-FRB associations at higher redshifts, specifically comparing the impact of two widely adopted Probability Density Functions (PDFs) for the dispersion measure.

\begin{figure}
\centering
\includegraphics[width=1.0\textwidth]{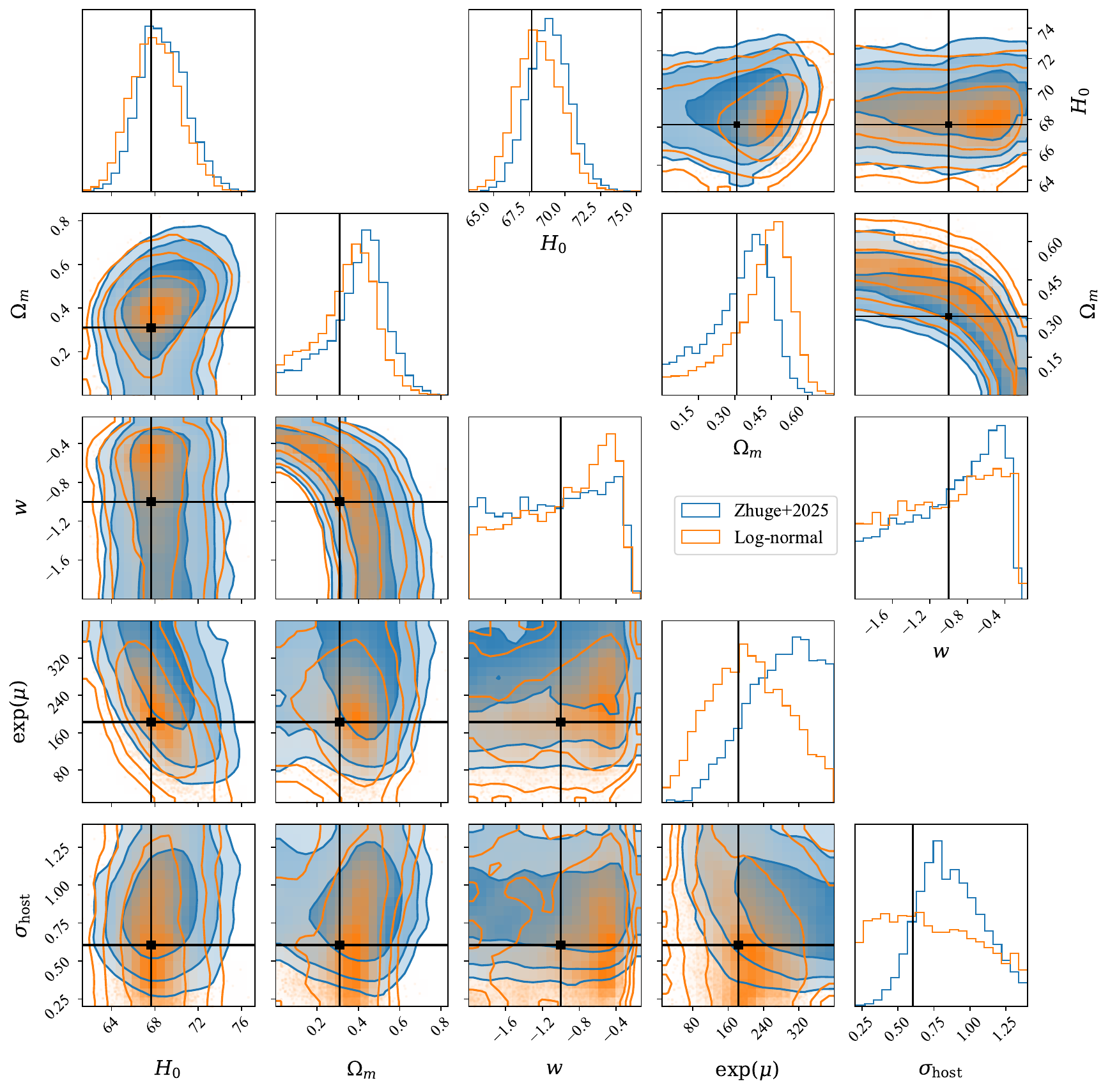}
\caption{Joint constraints on cosmological and host galaxy parameters for high-redshift case ($0.2<z<3.0$). The lower-left panels display the results obtained using the $(D_L, \DM_{\rm ext})$ dataset, while the upper-right panels show the constraints from the $(D_L, \DM_{\rm diff})$ dataset. The blue contours represent results obtained using Zhuge+2025 PDF \citep{Macquart_relation_2020}, as corrected by \citet{FRBcosmo_localised_Zhuge_et_al_2025}, while the orange contours correspond to the log-normal distribution. Solid black lines indicate the input values from \texttt{Planck18} \citep{Planck_Cosmo_param_2018}.}
\label{fig:MCMC_3}
\end{figure}

The idealized scenario, employing the $D_L-\DM_{\rm diff}$ data (where host galaxy contributions are excluded), is illustrated in the upper-right panels of Figure \ref{fig:MCMC_3}. Due to the extended redshift range, the constraints on $H_0$ and $\Omega_m$ are significantly improved compared to the low-redshift results. The $(H_0, \Omega_m)$ posterior contours are consistent with the idealized Gaussian case shown in Figure \ref{fig:idealised_3D_inference_and_marginalised}, yielding robust measurements for both parameters. Specifically, under the Zhuge+2025 PDF \citep{FRBcosmo_localised_Zhuge_et_al_2025}, we obtain $H_0 = \HighDiffMqH \text{ km s}^{-1} \text{ Mpc}^{-1}$, while the Log-normal distribution yields a consistent value of $H_0 = \HighDiffLnH \text{ km s}^{-1} \text{ Mpc}^{-1}$. However, the dark energy EoS $w$ remains poorly constrained ($w = \HighDiffMqw$ for Zhuge+2025 and $w=\HighDiffLnw$ for Log-normal), suggesting that $D_L-\DM_{\rm diff}$ data alone, even with CE-level precision, may struggle to distinguish between different $w\text{CDM}$ models due to uncertainties. Notably, the choice between the Zhuge+2025 and Log-normal PDFs does not lead to significant systematic shifts in the cosmological posteriors. This suggests that while each PDF has theoretical limitations, the cosmological inference is relatively robust to the specific modeling of the diffuse electron distribution.

Upon introducing the more realistic $D_L-\DM_{\rm ext}$ dataset — which accounts for both the diffuse and the host galaxy's electron density — the cosmological constraints remain remarkably stable, as evidenced in the lower-left panels of Figure \ref{fig:MCMC_3}. For instance, the Hubble constant is constrained to $H_0 = \HighExtMqH \text{ km s}^{-1} \text{ Mpc}^{-1}$ (Zhuge+2025) and $H_0 = \HighExtLnH \text{ km s}^{-1} \text{ Mpc}^{-1}$ (Log-normal). The interplay between the diffuse DM PDF and the host galaxy parameters ($\sigma_{\text{host}}$ and $\exp(\mu)$) presents an interesting secondary effect. We find that the median value of the host galaxy DM, $\exp(\mu)$, is better recovered when assuming a Log-normal distribution for the diffuse electron component ($\exp(\mu) = \HighExtLnmu \text{ pc cm}^{-3}$). Conversely, the parameter $\sigma_{\rm host}$ is more tightly constrained when the Zhuge+2025 PDF is utilized ($\sigma_{\rm host} = \HighExtMqsigma$). This indicates a degree of degeneracy between the modeling of the diffuse electron and the host environment over wide redshift ranges. Nevertheless, the vital conclusion is that these astrophysical uncertainties do not significantly bias the primary cosmological parameters. Our results demonstrate that even under the most realistic conditions and accounting for current observational uncertainties, the future Cosmic Explorer will be a powerful tool for the simultaneous study of global cosmology and the local environments of FRB host galaxies. Finally, our method provides a competitive alternative to other standard cosmology probes exploiting GWs with no direct redshift information from an EM counterpart in the context of future GWs detectors \citep[e.g.][]{Shiralilou_et_al_2023_DS_Forecast, Muttoni_et_al_2023_DS_Forecast, Cheng_Gair_2026_DS_Forecast}. They find a precision on $H_0$ of the level of $\lesssim 4 \%$ (till sub-percent levels), compared to our precision of about $3\%$, however they assume a network of two or more detectors, and mostly lower redshifts.

\begin{table}[ht]
    \centering
    \caption{Summary of MCMC constraints for high-redshift events ($0.2 < z < 3.0$)}
    \label{tab:MCMC_Highz}
    \begin{tabular}{lcccccc}
        \toprule
        Data & $\DM_{\rm diff}$ PDF Model & $H_0$ [$\text{km/s/Mpc}$] & $\Omega_m$ & $w$ & $\exp(\mu)$ [$\text{pc/cm}^3$] & $\sigma_{\rm host}$ \\
        \midrule
        $D_L-\DM_{\rm diff}$ & Zhuge+2025 & $\HighDiffMqH$ & $\HighDiffMqOmega$ & $\HighDiffMqw$ & --- & --- \\
        $D_L-\text{DM}_{\rm diff}$ & Log-normal & $\HighDiffLnH$ & $\HighDiffLnOmega$ & $\HighDiffLnw$ & --- & --- \\
        \addlinespace
        $D_L-\DM_{\rm ext}$  & Zhuge+2025 & $\HighExtMqH$ & $\HighExtMqOmega$ & $\HighExtMqw$ & $\HighExtMqmu$ & $\HighExtMqsigma$ \\
        $D_L-\DM_{\rm ext}$  & Log-normal & $\HighExtLnH$ & $\HighExtLnOmega$ & $\HighExtLnw$ & $\HighExtLnmu$ & $\HighExtLnsigma$ \\
        \bottomrule
    \end{tabular}
\end{table}

The quantitative results for all considered scenarios in the high-redshift regime are summarized in Table \ref{tab:MCMC_Highz}.

\section{Conclusions} \label{sec:conclusions}

In this work, we present a novel, redshift-independent framework for cosmological and astrophysical inference from the association between Fast Radio Bursts and Gravitational Waves. By utilizing the dispersion measure for FRBs and the luminosity distance for GWs, we demonstrate that it is possible to marginalize over redshift and constrain cosmological parameters. Our pipeline was validated both by using an idealized Gaussian distribution and subsequently tested with realistic mock events that incorporate current and future observational uncertainties.

Our analysis reveals a significant disparity between current and next-generation detector capabilities. As shown in Figure \ref{fig:MCMC_02}, while the current LIGO-Virgo network can constrain the Hubble constant $H_0$ within $2\sigma$, its performance in the low-redshift regime ($z < 0.2$) is primarily limited by sensitivity. In contrast, the Cosmic Explorer provides markedly more precise results, a direct consequence of its improved sensitivity \citep{2020JCAP_Jin_StandardSiren_CE, CE_sensitivity_2021}. 

Furthermore, compared to measurement uncertainty, the redshift range is a more essential factor when studying cosmology. We find that cosmological parameters constrained from low-redshift events are more easily affected by statistical fluctuations during data generation. Additionally, while these events are useful for $H_0$, they struggle to constrain $\Omega_m$ and $w$. This implies that the LV network requires a significantly larger number of events to achieve robust cosmological constraints. Conversely, the high-redshift results ($0.2<z < 3.0$) in Figure \ref{fig:MCMC_3} show that CE can successfully recover $(\Omega_m, H_0)$ with a posterior structure similar to the idealized Gaussian case (App.~\ref{App:Ideal_observations}). At the same time, $w$ remains poorly constrained due to its lower sensitivity in $E(z)$ at these distances. This indicates that even with next-generation detectors, GW-FRB associations, without redshift information, are most effective for studying $\Lambda$CDM and $H_0$ rather than complex dark energy dynamics.

Another central focus of this project was the impact of the host galaxy DM contribution. Given the potentially ``dirty'' environments of compact object mergers, $\DM_{\rm host}$ could differ significantly from that of standard FRB populations. Our strategy of treating host galaxy parameters ($\exp(\mu)$ and $\sigma_{\text{host}}$) as free parameters ensures that the pipeline remains robust, even if future observations reveal host environments with unexpected electron densities. Crucially, both the low-$z$ and high-$z$ analyses (Figures \ref{fig:MCMC_02} and \ref{fig:MCMC_3}) demonstrate that host galaxy properties do not significantly bias cosmological inference; they primarily increase the width of the constraints. This ``cosmological robustness'' is an encouraging sign for the utility of FRB-GW associations.

We further evaluated the systematic impact of the choice of PDFs for the diffuse electron DM. Despite the ongoing debate between Macquart's PDF \citep{Macquart_relation_2020, FRBcosmo_localised_Zhuge_et_al_2025} and Log-normal distributions \citep{Connor_2025NatAs, FRB_PDF_Sharma2025}, our results in Figure \ref{fig:MCMC_3} show that the specific choice of PDF does not lead to major shifts in cosmological parameters. The variations are largely confined to the host galaxy sector, suggesting that while cosmological simulations \citep[e.g.,][]{FRB_PDF_Alexander2024, FRB_PDF_Konietzka2025} continue to refine these models, our cosmological conclusions remain stable.

By simulating realistic NS-BH ``plunging'' events, we have demonstrated that accurate cosmology is achievable even in scenarios where the neutron star is swallowed without tidal disruption. While such events provide a ``clean'' environment for FRB propagation, they offer fewer electromagnetic tracers, making them a more challenging ``redshift-starved'' case. However, our results prove that even without direct redshift measurements, robust constraints can still be obtained. This underscores the strength of our methodology: if our framework succeeds in these conservative NS-BH scenarios, it will inherently yield even more precise results for NS-NS mergers, where electromagnetic counterparts are more likely to provide explicit redshift and inclination information. As we move toward the era of next-generation detectors, if GW-FRB associations are real, they are expected to be discovered in significant numbers, becoming a cornerstone of multi-messenger cosmology and providing a vital independent cross-check for both standard siren and standard candle measurements.

\begin{acknowledgments}

We thank Jin-Ping Zhu for helpful suggestions regarding the merger rate models. 
JZ's work is supported by a Top Tier Doctoral Graduate Research Assistantship (TTDGRA) at the University of Nevada, Las Vegas. 
JZ, CJH and BZ's work is supported by the Nevada Center for Astrophysics, NASA 80NSSC23M0104. 
CJH also acknowledges the support from the National Science Foundation through Award~PHY-2409727.
The authors are grateful for computational resources provided by the LIGO Laboratory and supported by NSF Grants PHY-0757058 and PHY-0823459.
This work also used the Expanse supercomputer at the San Diego Supercomputer Center (SDSC) through an allocation from the Advanced Cyberinfrastructure Coordination Ecosystem: Services \& Support (ACCESS) program (supported by NSF grants \#2138259, \#2138286, \#2138307, \#2137603, and \#2138296). We acknowledge \citet{Expanse} for providing the detailed technical architecture of the Expanse system.

\end{acknowledgments}

\vspace{5mm}

\emph{\large Software}: {\verb|emcee| \citep{emcee}, 
\verb|Astropy| \citep{Astropy_2013, Astropy_2018}, 
\verb|Numpy| \citep{2020_Numpy}, 
\verb|Scipy| \citep{2020_SciPy}, 
\verb|Matplotlib| \citep{2007_Matplotlib}, 
\verb|Bilby| \citep{Bilby_paper_2019, Bilby_cbc_2020}, 
\verb|corner| \citep{corner}.} 
The core pipeline and reproduction codes for the cosmological analysis in this work are publicly available on GitHub (\url{https://github.com/MariosNT/FRB_GW_association_cosmo}) and permanently archived on Zenodo via \dataset[doi:10.5281/zenodo.20282512]{https://doi.org/10.5281/zenodo.20282512} \citep{Code_zenodo}.

\appendix

\section{$H_0$ scaling relation at the low-z limit}\label{App:Low_z_limit}

We report here a useful scaling relation for $H_0$ as a function of $\DM_{\rm diff}$ and $D_L$ for small redshifts ($z \ll 1$). In such case, both expressions for the dispersion measure, eq. (\ref{eq:DM_diff}), and the luminosity distance, eq. (\ref{eq:luminosity_distance}), get simplified, and we can derive a simple estimate for the Hubble-Lemaitre parameter $H_0$. More specifically, we have
\begin{equation}
    D_L = \frac{cz}{H_0},
\end{equation}
for the luminosity distance, and
\begin{equation}
    \langle \DM_{\rm diff} \rangle = \frac{21}{64} \frac{c \Omega_b f_d}{\pi G m_p} H_0 z,
\end{equation}
for the diffuse electron dispersion measure. Both are linearly dependent on redshift and one can easily solve for $H_0$ to get
\begin{equation}
    H_0 = \sqrt{\frac{\langle \DM_{\rm diff} \rangle}{D_L} \frac{64}{21} \frac{\pi G m_p}{\Omega_b f_d}}.
\end{equation}
Normalizing the two distance measures at $z=0.01$ and assuming a fiducial, flat $\Lambda$CDM cosmology, we have

\begin{equation}
    H_0 = 67.49\ \frac{{\rm km}}{{\rm s \cdot Mpc}} \cdot \sqrt{\frac{\langle \DM_{\rm diff} \rangle}{8.22 \frac{{\rm pc}}{\rm cm^3}} \cdot \frac{44.65\ {\rm Mpc}}{D_L}}.
\end{equation}

Note that since this result applies at low redshifts, we assume relative precise values for $D_L$ and $\DM_{\rm diff}$, i.e. their PDFs can be modelled as very tight Gaussians. For an extension of this method, when the distance measures are very precise, but for arbitrary redshifts, we refer to App.~\ref{App:Ideal_observations}.

\section{Mock FRB and GWs catalogues}\label{App:mock_frb_gws_catalogues}

We include here more details about the different merger rate distributions of NS-BH events considered in the main text (see section \ref{sec:Redshift_distribution}). The three models we take from \cite{NSBH_merger_rates_Zhu_et_al_2021} are formulated as a redshift distribution $f(z)$ \citep{fz_factor_Sun_Zhang_Li_2015} multiplied by the current merger rate $\dot\rho_0$, $\dot\rho(z) = \dot \rho_0 f(z)$. The different models are:

\begin{itemize}
    \item \emph{Gaussian delay model} \citep{Rates_Gaussian_Virgilli_et_al_2011}:

    \begin{align}
        f_G(z) &= \Bigg[ (1+z)^{3.879 \eta} +\left( \frac{1+z}{73.5}\right)^{-0.4901 \eta} + \left( \frac{1+z}{3.672}\right)^{-5.691 \eta} + \\ \nonumber 
        &\left( \frac{1+z}{3.411}\right)^{-11.46 \eta} +\left( \frac{1+z}{3.546}\right)^{-16.38 \eta} + \left( \frac{1+z}{3.716}\right)^{-20.66 \eta}  \Bigg]^{1/\eta},
    \end{align}
    where $\eta=-7.553$.

    \item \emph{Log-normal delay model} \citep{Rates_LogNormal_PowerLaw_Wanderman_Piran_2015}:

    \begin{align}
        f_{\rm LN}(z) &= \Bigg[ (1+z)^{4.131 \eta} +\left( \frac{1+z}{22.37}\right)^{-0.5789 \eta} + \left( \frac{1+z}{2.978}\right)^{-4.735 \eta} + \\ \nonumber 
        &\left( \frac{1+z}{2.749}\right)^{-10.77 \eta} +\left( \frac{1+z}{2.867}\right)^{-17.51 \eta} + \left( \frac{1+z}{3.04}\right)^{-(0.08148+z^{0.574}/0.08682)\eta}  \Bigg]^{1/\eta},
    \end{align}
    where $\eta=-5.51$.

    \item \emph{Power Law delay model} \citep{Rates_LogNormal_PowerLaw_Wanderman_Piran_2015}:

    \begin{equation}\label{eq:app_power_law_redshift}
        f_{\rm PL}(z) = \Bigg[ (1+z)^{1.895 \eta} +\left( \frac{1+z}{5.722}\right)^{-3.759 \eta} + \left( \frac{1+z}{11.55}\right)^{-0.7426 \eta} \Bigg]^{1/\eta},
    \end{equation}
    where $\eta=-8.161$. This is the fiducial model in this work.
    
\end{itemize}

All three models are empirical fits on simulated data of merger rates from $z=0$ to $z=8$. On top of these models, we test a uniform redshift distribution, and also the empirical formula \citep{Rate_distribution_Zhao_et_al_2011, Rate_distribution_Cai_Yang_2017} for BNS systems:
\begin{equation}\label{eq:simple_merger_rate}
    P(z) \sim \frac{4 \pi D_c^2(z)}{H(z)} \frac{R(z)}{1+z},
\end{equation}
where $D_c(z)$ is the comoving distance and $R(z)$ is the evolving merger rate \citep{Rate_redshift_Schneider_et_al_2001, Rate_redshift_Cutler_Holz_2009, Rate_distribution_Cai_Yang_2017} given by:

\begin{equation}
    R(z) = \begin{cases}
1+2z, \quad z\leq 1\\
\frac{3}{4}(5-z), \quad 1<z<5\\
0, \quad z\geq 5
\end{cases}
\end{equation}

\begin{figure}[h]
\centering
\includegraphics[width=\textwidth]{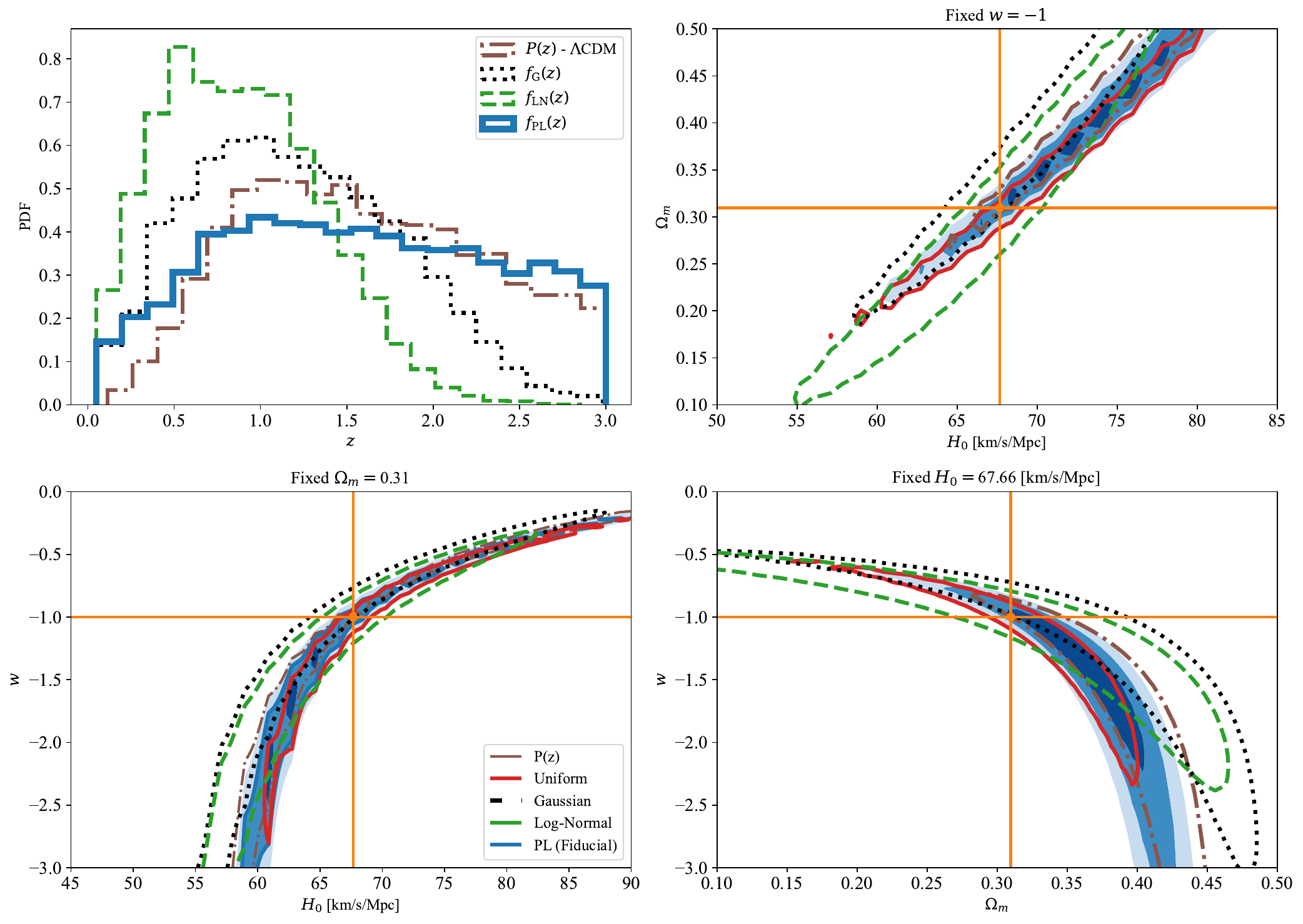}
\caption{(\emph{Top Left}) Comparison of four different merger rate models: Gaussian, Log-Normal, Power-Law, and the BNS empirical formula, Eq. (\ref{eq:simple_merger_rate}). Apart from the Log-normal model which peaks at lower redshifts ($z \sim 0.5$), the other models are very similar. (\emph{Contours}) 3D Inference after fixing one of the three cosmological parameters. Similar with left column of Figure \ref{fig:idealised_3D_inference_and_marginalised}, but comparing the different data generation models described in Section \ref{sec:Redshift_distribution}. We use $N_{\rm events}=50$ events. For the fiducial model, shades represent $68\%, 95\%$ and $99\%$ probability contours, while for the other models we only plot the $95 \%$ contour for better visualisation.}
\label{fig:idealised_3D_inference_multiple_rates}
\end{figure}

We compare the redshift distribution of the three different models, together with the BNS model $P(z)$, Eq.~\eqref{eq:simple_merger_rate}, in the top-left plot of Figure \ref{fig:idealised_3D_inference_multiple_rates}. We show that for our range of interest, all models are equivalent, with only the ``Log-normal'' one peaking at lower redshifts. We assess their impact on cosmological constraints in App. ~\ref{App:Ideal_observations}, finding negligible differences on the inference (see the other panels of the same Figure). As such, the fiducial choice of the ``power-law'' model is not influencing the results for the number of events and detector sensitivity considered in this work.

\section{Ideal observations}\label{App:Ideal_observations}

In this section, we consider the case where the errors in either $D_L$, or ${\rm DM}_{\rm diff}$, or both, become significantly small. For both of them, we assume a Gaussian distribution and perform a $\chi^2$ analysis for cosmological inference. This corresponds to a highly optimistic scenario, and provides a framework to study the impact of the different redshift distribution models in the final results, as well as the robustness of our Bayesian methodology in the main text.

Without loss of generality, we assume here a very precise $D_L$ determination. Since the luminosity distance is a monotonous function of redshift for a given cosmology, for each choice of cosmological parameters ($H_0, \Omega_m, w$) one has a 1-1 correspondence between $D_L$ and $z$ by solving $D_L = f(z, H_0, \Omega_m, w)$ from Section \ref{sec:dl_from_GWs}:
\begin{equation}
    D_L^{\rm obs}\quad \underrightarrow{(H_0, \Omega_m, w)}\quad z.
\end{equation}

At the same time, there is a connection between redshift and the dispersion measure of an FRB, as developed in Section \ref{sec:dm}. Hence, for a given cosmology\footnote{In reality, additional astrophysical parameters influence this relationship, like the fraction of the diffuse baryons $f_{\rm diff}$. Unless stated otherwise, we fix all other parameters to their fiducial values, as stated in the text.}, one can predict a theoretical $\DM_{\rm diff}$:
\begin{equation}
    z \quad \underrightarrow{(H_0, \Omega_m, w)}\quad {\rm DM_{\rm diff}}.
\end{equation}

\begin{figure}
\centering
\includegraphics[width=\textwidth]{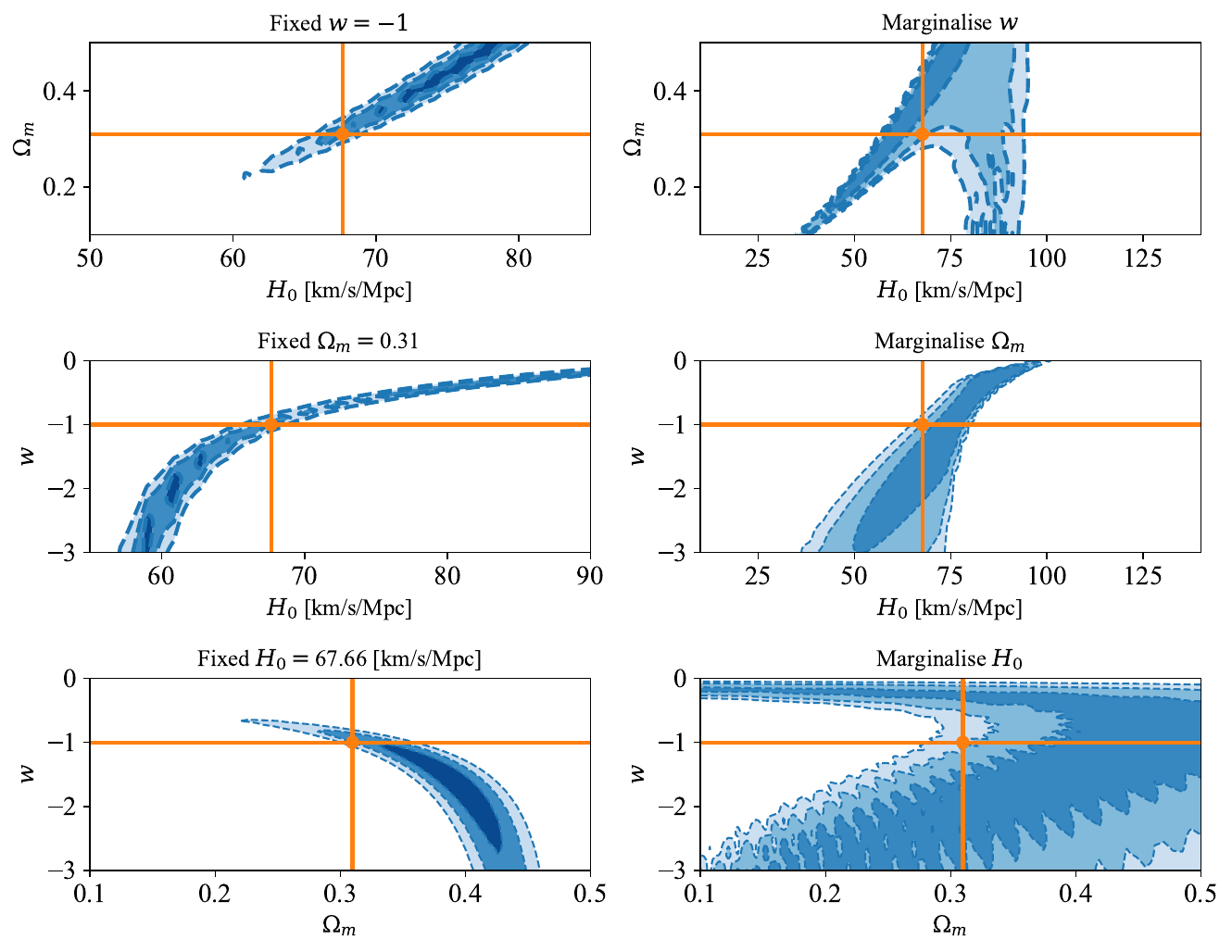}
\caption{3D Inference for the fiducial data generation model, for $N_{\rm events}=50$ events, based on ideal observations. Shades represent $68\%, 95\%$ and $99\%$ probability contours. Left column shows results after fixing one of the three cosmological parameters. Right column shows results after marginalising over the parameter not shown.}
\label{fig:idealised_3D_inference_and_marginalised}
\end{figure}

The latter can then be compared to ${\DM}_{\rm diff}^{\rm obs}$, and cosmological constraints are set with the standard $\chi^2$ statistic:

$$
\chi^2 (H_0, \Omega_m, w) = \frac{[{\rm DM}_{\rm diff}^{\rm obs}-{\rm DM}_{\rm diff}(H_0, \Omega_m, w)]^2}{\sigma_{{\rm DM_{\rm diff}}}^2},
$$

where a $3D$ grid is used for the triplet ($H_0, \Omega_m, w$) and for $\sigma_{{\rm DM_{\rm diff}}}$ we adopt a value of $\sigma_{{\rm DM_{\rm diff}}} \sim 105$ pc/cm$^3$ \citep{Wei_et_al_FRB_GWs_Ass_2018}. The ranges of the different parameters are $H_0 \ \epsilon \ [10, 140]$ km/s/Mpc, $\Omega_m \ \epsilon \ [0.1, 0.5]$ and $w \ \epsilon \ [-3, 0]$. The simulation of a population of GWs and FRB events which provides the pair of ($D_L, \DM_{\rm diff}$) is the one described in Section \ref{sec:simulated_data}. For this test study we generate $N_{\rm events}=50$ events and we perform the following analyses: 

\begin{enumerate}
    \item In Figure \ref{fig:idealised_3D_inference_multiple_rates} study the impact of using different redshift distributions for our events (Section \ref{sec:Redshift_distribution}, App.~\ref{App:mock_frb_gws_catalogues}). We show only the 2D constraints after fixing the third parameter, since these provide the most optimistic cases. We show that based on current capabilities, it is not possible to distinguish between the different models. As such the modeling choice made in Section \ref{sec:Redshift_distribution}, the events following the power-law delay redshift model - is not expected to influence our conclusions.
    
    \item In Figure \ref{fig:idealised_3D_inference_and_marginalised}, we repeat the analysis of Figure \ref{fig:idealised_3D_inference_multiple_rates}, under the fiducial power-law delay model. We perform a 3D inference of $(H_0, \Omega_m, w)$ and show the relevant 2D contours in each case, after fixing or marginalizing over the missing parameter. In all cases we are recovering the input values, with the marginalized contours being larger as expected. The results in the left column can be compared to Figure \ref{fig:idealised_3D_inference_multiple_rates}, where the blue contours show again a different realisation of the power-law delay model.
\end{enumerate}

The idealized results demonstrate the validity of the method, and confirm that even at an optimistic scenario the small differences between the redshift distribution models (App.~\ref{App:mock_frb_gws_catalogues}) are not impacting the cosmological inference in a distinguishable way.

\section{Viewing Angle Effects}\label{App:viewing_angle}

In the main text, Section \ref{sec:GW_gen}, we mentioned that all our NS-BH sources were simulated with a fixed input viewing angle\footnote{In the case of non-spinning compact objects, as in our case, this angle is equivalent to the inclination of the orbit $i$. Recall Section \ref{sec:dl_from_GWs}.}: $\theta_{\rm JN} = 0.4$, while in our parameter inference we assumed no informative prior information or constraints in the angle. In this section we discuss the impact of these choices in our modelling and cosmological constraints. We investigate two questions:
\begin{enumerate}
    \item What is the impact of simulating NS-BH GWs events with a variety of input angles?
    \item What is the impact of having prior information that constrains $\theta_{\rm JN}$?
\end{enumerate}

\begin{figure}[h]
\centering
\includegraphics[width=\textwidth]{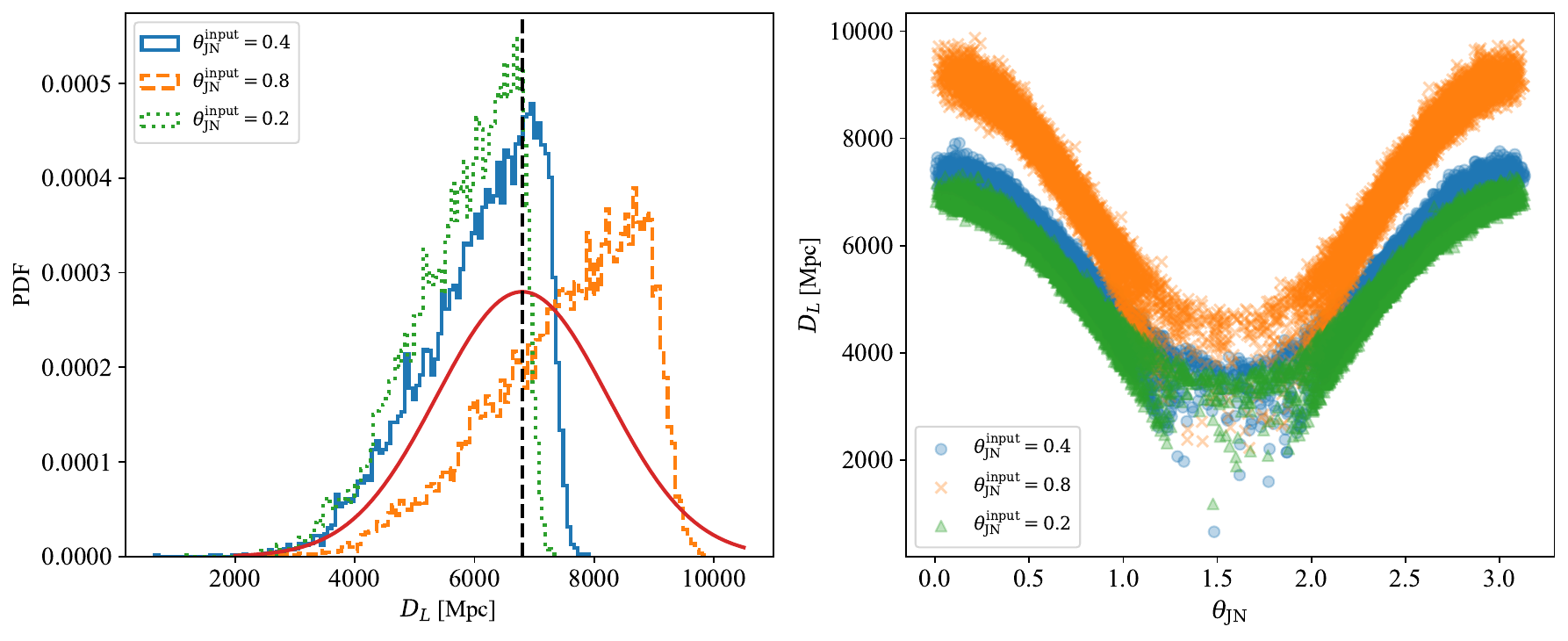}
\caption{Investigating the impact of different input viewing angles. Recall that $\theta_{\rm JN}^{\rm input}=0.4$ is the fiducial choice used in the main text. (\emph{Left}) Luminosity distance posterior histograms for each choice of input viewing angle, which correspond to the 1D marginalised distributions of the samples shown on the right figure. Larger $\theta_{\rm JN}^{\rm input}$ leads to a signal with smaller amplitude, which the parameter inference mainly assigns to a larger possible maximum distance. The red, solid line corresponds to the Gaussian used to generate $D_L$ values for our simulated events (section \ref{sec:simulated_data}). (\emph{Right}): 2D posterior samples in the luminosity distance - viewing angle parameter space. We clearly observe the degeneracy of the two parameters (section \ref{sec:dl_from_GWs}), which is not improved for any choice of $\theta_{\rm JN}$. See text for more details.}
\label{fig:input_angle_effects}
\end{figure}

To answer the first question we simulate two additional GWs events, detected by Cosmic Explorer at $z=1$, but with different input viewing angles $\theta_{\rm JN}^{\rm input}$. We keep all other parameters the same, while for the angles we choose an approximate ``face-on'' case ($\theta_{\rm JN}^{\rm input} =0.2$) and an approximate ``edge-on'' case ($\theta_{\rm JN}^{\rm input}=0.8$). We show our results in Figure \ref{fig:input_angle_effects}. In all cases, we recover the other physical parameters of the binary, e.g. masses, very accurately. As expected, in the ``edge-on'' case, the amplitude of the signal is weaker, so the inference allows for larger distances, while for smaller angles the strength of the signal restricts the inference to smaller distances. Without any prior information on the viewing angle, the well-known degeneracy between $D_L-\theta_{\rm JN}$ appears in all cases. More importantly, with this test we wanted to confirm that a single choice of $\theta_{\rm JN}^{\rm input}$ does not induce any bias to our results: Indeed, we verify that this is the case overplotting the Gaussian distribution used for sampling the ``observed'' $D_L$ for an event at $z=1$ (recall Section \ref{sec:generating_mock_events}). For a specific redshift, we draw a random sample from a Gaussian centered at the fiducial $\Lambda$CDM $D_L$ for this redshift, with a width calculated by realistic GWs simulations (Section \ref{sec:GW_gen} and Figure \ref{fig:DL_posteriors_LV_CE}). In Figure \ref{fig:input_angle_effects} we observe that this distribution covers the full range of distances inferred for possible input viewing angles, so no bias is introducing by fixing $\theta_{\rm JN}^{\rm input}=0.4$ for our simulated events.

\begin{figure}[h]
\centering
\includegraphics[width=\textwidth]{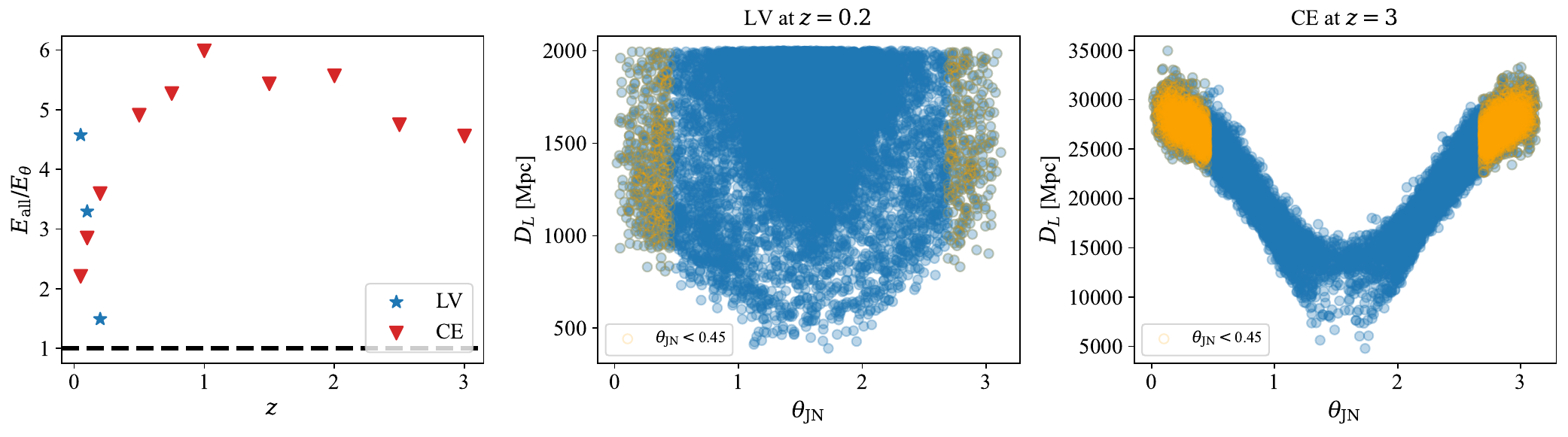}
\caption{Investigating the impact of viewing angles constraints on the luminosity distance estimates. (\emph{Left}) Improvement on the fractional error $\Delta_{D_L}/D_L$ for a range of redshifts and different detectors, when angle prior information is present $E_{\theta}$ vs when it is not $E_{\rm all}$. (\emph{Middle} $\&$ \emph{Right}) 2D posterior samples in the luminosity distance - viewing angle parameter space for LV (middle) and CE (right), for the largest redshift, i.e. worst case scenario, for each. We observe that for LV the $D_L$ inference is so wide, that even when restricting the angles one does not gain a significant insight - this leads to the declining trend in the left panel. For CE, the samples are still quite tight, so constraining the viewing angle leads to meaningful improvement.}
\label{fig:error_improvement_on_theta_constraints}
\end{figure}

To answer the second question we impose an artificial cut on the posterior samples we have, based on the viewing angle. More specifically, we impose a constraint on the angles to be $\theta_{\rm JN}<0.45$ or $\theta_{\rm JN}>\pi-0.45$. This choice corresponds to an approximate ``face-on'' orbit, and in a realistic scenario it can be the outcome of additional constraints coming for a coincident Gamma-Ray Burst (GRB) jet observation. The main effect of restricting the possible angles is the break of the $D_L-i$ degeneracy, and as a result a better determination of $D_L$. In Figure \ref{fig:error_improvement_on_theta_constraints}, we show the improvement of the relative luminosity distance error for different redshifts and detectors when we have prior information that limits the range of the viewing angle. Again, we have calculated the luminosity distance error as described in Section \ref{sec:GW_gen}. We show that: 1) in all cases, the ratio of the error before $E_{\rm all}$ and after $E_\theta$ the angle constraints is bigger than one, i.e. the distance posteriors are tightened as expected. 2) For a CE detector, the improvement reaches a limit between $5$ and $6$ after $z=0.75$ which results from the relative constancy of the fractional distance error $\Delta_{D_L}/D_L$ after some redshift, i.e. both terms in the fraction get larger at a similar rate (see right panel of Figure \ref{fig:error_improvement_on_theta_constraints}). 3) In contrast, for a LV network we see that the improvement worsens with redshift. This is a consequence of the generally much weaker constraints on $D_L$ for larger redshifts, where the extra $\theta_{\rm JN}$ constraints cannot provide any additional information (see middle panel of Figure \ref{fig:error_improvement_on_theta_constraints}). In summary, as anticipated, additional knowledge about the viewing angle is improving the distance inference, and as a consequence it is expected to ameliorate the cosmological constraints. This result is consistent with previous analysis that investigate the systematic effects of an accurate determination of the viewing angle in constraints of the Hubble parameter \citep{Chen_GWs_angle_2020, Salvarese_Chen_GWs_angle_2024}.

\vspace{5mm}

\newpage
\bibliography{ref}{}
\bibliographystyle{aasjournal}
\end{CJK}
\end{document}